\documentclass[pdflatex,sn-mathphys-num]{sn-jnl}

\usepackage{algorithm}
\usepackage{algorithmicx}
\usepackage{algpseudocode}
\usepackage{amsfonts}
\usepackage{amsmath}
\usepackage{amssymb}
\usepackage{amsthm}
\usepackage[title]{appendix}
\usepackage{blkarray}
\usepackage{booktabs}
\usepackage{caption}
\usepackage{cleveref}
\usepackage{comment}
\usepackage{float}
\usepackage{graphicx}
\usepackage{hyperref}
\usepackage{listings}
\usepackage{manyfoot}
\usepackage[scr=boondox,cal=esstix]{mathalpha}
\usepackage{mathrsfs}
\usepackage{multirow}
\usepackage{ragged2e}
\usepackage{subfig}
\usepackage{tabularx}
\usepackage{textcomp}
\usepackage{verbatim}
\usepackage{xcolor}
\usepackage{notes2bib}
\usepackage{soul}

\renewcommand\parallel{\mathbin{/\!/}}

\hypersetup{colorlinks=true, citecolor=orange, urlcolor=blue, linkcolor=magenta}

\begin{document}

\title[Article Title]{Entropy-based models to randomize real-world hypergraphs}

\author*[1,2,3]{\fnm{Fabio} \sur{Saracco}}\email{fabio.saracco@cref.it}

\author[4]{\fnm{Giovanni} \sur{Petri}}\email{giovanni.petri@nulondon.ac.uk}

\author[5]{\fnm{Renaud} \sur{Lambiotte}}\email{renaud.lambiotte@maths.ox.ac.uk}

\author[3,6]{\fnm{Tiziano} \sur{Squartini}}\email{tiziano.squartini@imtlucca.it}

\affil*[1]{\orgname{`Enrico Fermi' Research Center (CREF)}, \orgaddress{\street{Via Panisperna 89A}, \city{Rome}, \postcode{00184}, \country{Italy}}}

\affil[2]{\textcolor{black}{\orgdiv{`Mauro Picone' Institute for Applied Computing (IAC)}, \orgname{CNR}, \orgaddress{\street{Via Madonna del Piano 10}, \city{Sesto Fiorentino}, \postcode{50019}, \country{Italy}}}}

\affil[3]{\orgname{IMT School for Advanced Studies}, \orgaddress{\street{Piazza San Francesco 19}, \city{Lucca}, \postcode{55100}, \country{Italy}}}

\affil[4]{\orgdiv{Network Science Institute}, \orgname{Northeastern University London}, \orgaddress{\street{Devon House 58 St. Katharine's Way}, \city{London}, \postcode{E1W 1LP}, \country{United Kingdom}}}

\affil[5]{\orgdiv{Mathematical Institute}, \orgname{University of Oxford}, \orgaddress{\street{Woodstock Road}, \city{Oxford}, \postcode{OX2 6GG}, \country{United Kingdom}}}

\affil[6]{\orgdiv{INdAM-GNAMPA}, \orgname{Istituto Nazionale di Alta Matematica `Francesco Severi'}, \orgaddress{\street{P.le Aldo Moro 5}, \city{Rome}, \postcode{00185}, \country{Italy}}}

\abstract{Network theory has often disregarded many-body relationships, solely focusing on pairwise interactions: neglecting them, however, can lead to misleading representations of complex systems. Hypergraphs represent a suitable framework for describing polyadic interactions. Here, we leverage the representation of hypergraphs based on the incidence matrix for extending the entropy-based approach to higher-order structures: in analogy with the Exponential Random Graphs, we introduce the Exponential Random Hypergraphs (ERHs). After exploring the asymptotic behaviour of thresholds generalising the percolation one, we apply ERHs to study real-world data. First, we generalise key network metrics to hypergraphs; then, we compute their expected value and compare it with the empirical one, in order to detect deviations from random behaviours. Our method is analytically tractable, scalable and capable of revealing structural patterns of real-world hypergraphs that differ significantly from those emerging as a consequence of simpler constraints.}



\maketitle

\section*{Introduction}\label{sec:intro}

Networks provide a powerful language to model interacting systems~\cite{Newman2010,Caldarelli2010}. Within such a framework, the basic unit of interaction, i.e. the \emph{edge}, involves two nodes and the complexity of the  structure as a whole arises from the combination of these units. Despite its many successes, network science disregards certain aspects of interacting systems, notably the possibility that more-than-two constituent units could interact at a time~\cite{Lambiotte2019}. Yet, it has been increasingly shown that, for a variety of systems, interactions cannot be always decomposed into a pairwise fashion and that neglecting higher-order ones can lead to an incomplete, if not misleading, representation of them~\cite{Lambiotte2019,Battiston2020,bick2023higher}: examples include chemical reactions involving several compounds, coordination activities within small teams of co-working people and brain activities mediated by groups of neurons. Generally speaking, thus, modelling the joint coordination of multiple entities calls for a generalisation of the traditional edge-centred framework.

\textcolor{black}{While approaches focusing on the so-called \emph{simplicial complexes} have been proposed~\cite{courtney2016generalized,bianconi2022statistical}, an increasingly popular alternative to support a science of many-body interactions is provided by \emph{hypergraphs}, as these mathematical objects allow nodes to interact in groups without posing restrictions such as the `hierarchical' ones characterising the former ones~\cite{Battiston2021} - which, in fact, include all the subsets of a given simplex~\cite{courtney2016generalized}.}

\textcolor{black}{Several contributions to the definition of analytical tools for their study have already appeared~\cite{Chodrow2020,Nakajima2021,Musciotto2021,Barthelemy2022}: while some pertain to the purely mathematical literature and have considered probabilistic hypergraphs with the aim of studying properties such as the existence of cycles, cliques, etc.~\cite{Dudek2017,Adser2019,Barrett2024}, others have adopted approaches rooted into statistical physics. Among the latter ones, some have proposed microcanonical approaches~\cite{Chodrow2020,Musciotto2021} while others have focused on canonical ones~\cite{Nakajima2021,wegner2021atomic,Barthelemy2022}.}

\textcolor{black}{The present contribution aims at extending the class of entropy-based benchmarks~\cite{Jaynes1957a,Park2004statistical,Garlaschelli2008,Squartini2011a} to hypergraphs while providing a coherent framework to formally derive the canonical approaches that have been proposed so far.} These models work by preserving a given set of quantities while randomising everything else, hence destroying all possible correlations between structural properties except for those that are genuinely embodied into the constraints themselves~\cite{Squartini2011a,Cimini2018a,Squartini2018}. The versatility of such an approach allows it to be employed either in presence of full information (to quantify the level of self-organisation of a given configuration by identifying the patterns that are incompatible with simpler, structural constraints~\cite{Squartini2013t,Saracco2017,Becatti2019d,Caldarelli2020b,Neal2021}) or in presence of partial information (to infer the missing portion of a given configuration~\cite{Parisi2020}).

Our strategy for defining null models for hypergraphs is based on the randomisation of their incidence matrix, i.e. the (generally, rectangular) table contains information about the connectivity of nodes (the set of hyperedges they belong to) and the connectivity of hyperedges (the set of nodes they cluster). We will explicitly derive two members of this novel class of models hereby named \emph{Exponential Random Hypergraphs}, i.e. the Random Hypergraph Model (RHM, generalising the Erd\"os-R\'enyi Model) and the Hypergraph Configuration Model (HCM, generalising the Configuration Model), and provide an analytical characterisation of their behaviour. To this aim, we will exploit the formal equivalence between the incidence matrix of a hypergraph and the biadjacency matrix of a bipartite graph~\cite{Nakajima2021,Barthelemy2022}. Afterwards, we will employ the HCM to assess the statistical significance of a number of patterns characterising several real-word hypergraphs.

\section*{Methods}\label{sec:meth}

\subsection*{Formalism and basic quantities}\label{ssec:formalism}

A hypergraph can be defined as a pair $H(\mathcal{V},\mathcal{E}_H)$ where $\mathcal{V}$ is the set of vertices and $\mathcal{E}_H$ is the set of hyperedges. Moving from the observation that the edge set $\mathcal{E}_G$ of a traditional, binary, undirected graph $G(\mathcal{V},\mathcal{E}_G)$ is a subset of the power set of $\mathcal{V}$, several definitions of the hyperedge set $\mathcal{E}_H$ have been provided: the two most popular ones are those proposed in~\cite{Berge1967,Berge1973}, where hyperedges tie one or more vertices, and in~\cite{Voloshin2009}, where hyperedges are allowed to be empty sets as well. Hereby, we adopt the definition according to which $\mathcal{E}_H$ is a multiset of the power set of $\mathcal{V}$: \textcolor{black}{since the concept of `multiset' extends the concept of `set', allowing for multiple instances of (each of) its elements, our choice implies that we are considering non-simple hypergraphs, admitting loops and parallel edges (i.e. hyperedges involving exactly the same nodes) of any size, including 0 (corresponding to empty hyperedges) and $|\mathcal{V}|$ (corresponding to hyperedges clustering all vertices together).}

As for traditional graphs, an algebraic representation of hypergraphs can be devised as well. In analogy with the traditional case, we call the cardinality of the set of nodes $|\mathcal{V}|\equiv N$ and the cardinality of the set of hyperedges $|\mathcal{E}_H|\equiv L$: then, we consider the $N\times L$ table known as \emph{incidence matrix}, each row of which corresponds to a node and each column of which corresponds to a hyperedge. If we indicate the incidence matrix with $\mathbf{I}$, its generic entry $I_{i\alpha}$ will be $1$ if vertex $i$ belongs to hyperedge $\alpha$ and $0$ otherwise. Notice that the number of $1$s along each row can vary between $0$ and $L$, the former case indicating an isolated node and the latter one indicating a node that belongs to each hyperedge; similarly, the number of $1$s along each column can vary between $0$ and $N$, the former case indicating an empty hyperedge and the latter one indicating a hyperedge that includes all nodes. As explicitly noticed elsewhere~\cite{Chodrow2020,Nakajima2021,Barthelemy2022}, representing a hypergraph via its incidence matrix is equivalent to considering the bipartite graph defined by the sets $\mathcal{V}$ and $\mathcal{E}_H$ \textcolor{black}{- more formally, the function that assigns each hypergraph to the bipartite graph associated with it is a bijection when both nodes and hyperedges are uniquely labeled~\cite{Chodrow2020}.} For instance, the incidence matrix $\mathbf{I}$ describing the binary, undirected hypergraph shown in Fig.~\ref{fig1} is the following:

\begin{figure}[t!]
\centering
\includegraphics[width=0.7\textwidth]{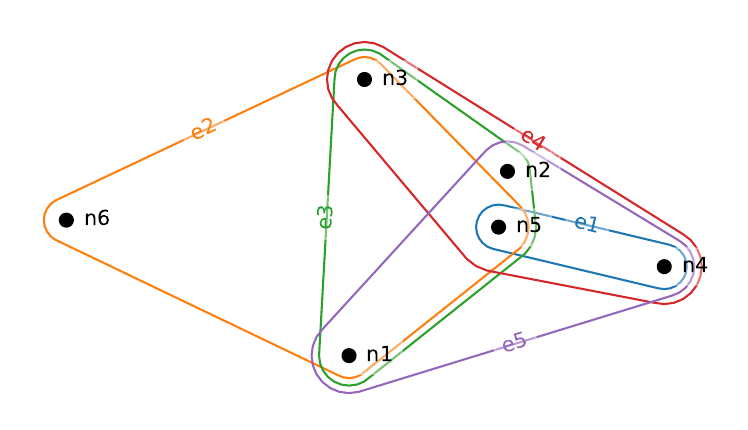}
\caption{\textcolor{black}{\textbf{Cartoon representation of a hypergraph.} Black dots represent nodes, while the colored shapes represent hyperedges. The incidence matrix describing the present hypergraph is defined by Eq.~\ref{eq:inc_mat}.}}
\label{fig1}
\end{figure}

\begin{equation}\label{eq:inc_mat}
\mathbf{I}=
\begin{blockarray}{cccccc}
& \mathbf{e_1} & \mathbf{e_2} & \mathbf{e_3} & \mathbf{e_4} & \mathbf{e_5} \\
\begin{block}{c(ccccc)}
  \mathbf{n_1} & 0 & 1 & 1 & 0 & 1 \\
  \mathbf{n_2} & 0 & 0 & 1 & 1 & 1 \\
  \mathbf{n_3} & 0 & 1 & 1 & 1 & 0 \\
  \mathbf{n_4} & 1 & 0 & 0 & 1 & 1 \\
  \mathbf{n_5} & 1 & 1 & 1 & 1 & 0 \\
  \mathbf{n_6} & 0 & 1 & 0 & 0 & 0 \\
\end{block}
\end{blockarray}\:.
\end{equation}

Once the incidence matrix has been defined, several quantities needed for the description of hypergraphs can be defined quite straightforwardly: for example, the `degree of node $i$' (hereby, \emph{degree}) reads 

\begin{equation}
k_i=\sum_{\alpha=1}^LI_{i\alpha}
\end{equation}
and counts the number of hyperedges that are incident to it; analogously, the `degree of hyperedge $\alpha$' (hereby, \emph{hyperdegree}) reads

\begin{equation}
h_\alpha=\sum_{i=1}^NI_{i\alpha}
\end{equation}
and counts the number of nodes it clusters. Both the sum of degrees and that of hyperdegrees equal the total number of $1$s, i.e. $\sum_{i=1}^Nk_i=\sum_{i=1}^N\sum_{\alpha=1}^LI_{i\alpha}=\sum_{\alpha=1}^L\sum_{i=1}^NI_{i\alpha}=\sum_{\alpha=1}^Lh_\alpha\equiv T$. Importantly, a node degree no longer coincides with the number of its neighbours: instead, it matches the number of hyperedges it belongs to; a hyperdegree, instead, provides information about the hyperedge size. \textcolor{black}{Analogously, $T$ paves the way for the alternative definition of `density of connections' reading $\rho=T/NL\equiv h/N$, i.e. the ratio between the (average) number of nodes each hyperedge clusters and the total number of nodes.}

\subsection*{Binary, undirected hypergraphs randomisation}

\textcolor{black}{An early attempt to define randomisation algorithms for hypergraphs within a framework closely resembling ours can be found in~\cite{Ghoshal2009a}.} Its authors, however, have just considered hyperedges that are incident to triples of nodes - a framework that has been, later, applied to the study of the World Trade Network~\cite{Nakajima2021}.

Considering the incidence matrix has two clear advantages over the tensor-based representation employed in~\cite{Ghoshal2009a,Nakajima2021}: \emph{i) generality}, because the incidence matrix allows hyperedges of any size to be handled at once; \emph{ii) compactness}, because the order of the tensor $\mathbf{I}$ never exceeds two, hence allowing any hypergraph to be represented as a traditional, bipartite graph.

In order to extend the rich set of null models induced by graph-specific global and local constraints to hypergraphs, we, first, need to identify the quantities that can play this role within the novel setting. In what follows, we will consider the total number of $1$s, i.e. $T$, the degree and the hyperdegree sequences, i.e. $\{k_i\}_{i=1}^N$ and $\{h_\alpha\}_{\alpha=1}^L$ - either separately or in a joint fashion; moreover, we will distinguish between microcanonical and canonical randomisation techniques.

\subsubsection*{Homogeneous benchmarks:\\the Random Hypergraph Model (RHM)}

\paragraph{Microcanonical formulation}

The model is defined by just one, global constraint, that, in our case, reads 

\begin{equation}
T=\sum_{i=1}^N\sum_{\alpha=1}^LI_{i\alpha};
\end{equation}
its microcanonical version extends the model by Erd\"os and R\'enyi~\cite{Erdos1959} - also known as Random Graph Model (RGM) - to hypergraphs and prescribes to count the number of incidence matrices that are compatible with a given, total number of $1$s, say $T^*$: they are

\begin{equation}
\Omega_\text{RHM}=\binom{V}{T^*}
\end{equation}
with $V\equiv NL$ being the total number of entries of the incidence matrix $\mathbf{I}$. Once the total number of configurations composing the microcanonical ensemble has been determined, a procedure to generate them is needed: in the case of the RHM, it simply boils down to reshuffling the entries of the incidence matrix, a procedure ensuring that the total number of $1$s is kept fixed while any, other correlation is destroyed.

\paragraph{Canonical formulation}

The canonical version of the RHM, instead, extends the model by Gilbert~\cite{Gilbert1959} and rests upon the constrained maximisation of Shannon entropy, i.e.

\begin{equation}
\mathscr{L}\equiv S[P]-\sum_{i=0}^M\theta_i[P(\mathbf{I})C_i(\mathbf{I})-\langle C_i\rangle]
\end{equation}
where $S[P]=-\sum_{\mathbf{I}\in\mathcal{I}}P(\mathbf{I})\ln P(\mathbf{I})$, $C_0\equiv\langle C_0\rangle\equiv 1$ sums up the normalisation condition and the remaining $M-1$ constraints represent proper topological properties. The sum defining Shannon entropy runs over the set $\mathcal{I}$ of incidence matrices described in the introductory paragraph and known as \textit{canonical ensemble}. Such an optimisation procedure defines the \textit{Exponential Random Hypergraph} (ERH) framework, described by the expression

\begin{equation}
P(\mathbf{I})=\frac{e^{-H(\mathbf{I})}}{Z}=\frac{e^{-H(\mathbf{I})}}{\sum_{\mathbf{I}\in\mathcal{I}}e^{-H(\mathbf{I})}}=\frac{e^{-\sum_{i=1}^M\theta_iC_i(\mathbf{I})}}{\sum_{\mathbf{I}\in\mathcal{I}}e^{-\sum_{i=1}^M\theta_iC_i(\mathbf{I})}}.
\end{equation}

In the simplest case, the only global constraint is represented by $T$ and leads to the expression

\begin{align}
P(\mathbf{I})&=\frac{e^{-\theta T(\mathbf{I})}}{\sum_{\mathbf{I}\in\mathcal{I}}e^{-\theta T(\mathbf{I})}}=\frac{e^{-\sum_{i=1}^N\sum_{\alpha=1}^L\theta I_{i\alpha}}}{\sum_{\mathbf{I}\in\mathcal{I}}e^{-\sum_{i=1}^N\sum_{\alpha=1}^L\theta I_{i\alpha}}}=\prod_{i=1}^N\prod_{\alpha=1}^L x^{I_{i\alpha}}\prod_{i=1}^N\prod_{\alpha=1}^L(1+x)^{-1}
\end{align}
that can be re-written as

\begin{align}
P(\mathbf{I})=\prod_{i=1}^N\prod_{\alpha=1}^Lp^{I_{i\alpha}}(1-p)^{1-I_{i\alpha}}=p^{T(\mathbf{I})}(1-p)^{NL-T(\mathbf{I})}
\end{align}
with $e^{-\theta}\equiv x$ and $p\equiv x/(1+x)$. The canonical ensemble, now, includes all $N\times L$, rectangular matrices whose number of entries equal to 1 ranges from 0 to $NL$. According to such a model, the entries of the incidence matrix are i.i.d. Bernoulli random variables, i.e. $I_{i\alpha}\sim\text{Ber}(p)$, $\forall\:i,\alpha$; as a consequence, the total number of $1$s, the degrees and the hyperdegrees obey Binomial distributions, being all defined as sums of i.i.d. Bernoulli random variables: specifically, $T\sim\text{Bin}(NL,p)$, $k_i\sim\text{Bin}(L,p)$, $\forall\:i$ and $h_\alpha\sim\text{Bin}(N,p)$, $\forall\:\alpha$, in turn, implying that $\langle T\rangle_\text{RHM}=NLp$, $\langle k_i\rangle_\text{RHM}=Lp$, $\forall\:i$ and $\langle h_\alpha\rangle_\text{RHM}=Np$, $\forall\:\alpha$.

\paragraph{Parameter estimation}

In order to ensure that $\langle T\rangle_\text{RHM}=T^*$, parameters have to be tuned opportunely. To this aim, the likelihood maximisation principle can be invoked~\cite{Garlaschelli2008}: it prescribes to maximise the function $\mathcal{L}(\theta)\equiv\ln P(\mathbf{I}^*|\theta)$ with respect to the unknown parameter that defines it. Such a recipe leads us to find 

\begin{equation}
p=\rho^*=\frac{T^*}{NL}
\end{equation}
with $T^*=T(\mathbf{I}^*)$ indicating the empirical value of the constraint defining the RHM.\\

\textcolor{black}{The RHM (also considered in~\cite{Barthelemy2022}, although without providing any derivation from first principles, and in~\cite{wegner2021atomic}, although without providing any recipe for the estimation of its parameter)} is formally equivalent to the Bipartite Random Graph Model (BiRGM)~\cite{Saracco2015}. Such an identification is guaranteed by our focus on non-simple hypergraphs.

\paragraph{Estimation of the number of empty hyperedges}

\textcolor{black}{Non-simple hypergraphs admit the presence of empty as well as parallel hyperedges. As this type of structures is associated with configurations that may be regarded as problematic (since not observed in empirical data), we evaluate how frequently they appear in the ensembles induced by our benchmarks. Let us denote the number of empty hyperedges, i.e. the number of hyperedges whose hyperdegree equals zero, with $N_\emptyset$: since the hyperdegrees are i.i.d. Binomial random variables, $N_\emptyset\sim\text{Bin}(L,p_\emptyset)$ where}

\begin{equation}
\textcolor{black}{p_\emptyset\equiv (1-p)^N}
\end{equation}
\textcolor{black}{is the probability for the generic hyperedge to be empty or, equivalently, for its hyperdegree to equal zero; the expected number of empty hyperedges reads}

\begin{equation}
\textcolor{black}{\langle N_\emptyset\rangle=Lp_\emptyset=L(1-p)^N.}
\end{equation}

Let us, now, inspect the behaviour of $p_\emptyset$ on the ensemble induced by the RHM as the density of $1$s in the incidence matrix, i.e. $p=T/NL$, varies. The two regimes of interest are the \emph{dense} one, defined by $T\rightarrow NL$, and the \emph{sparse} one, defined by $T\rightarrow0$. In the dense case, one finds

\begin{align}
\lim_{T\to NL}p_\emptyset&=\lim_{T\to NL}\left(1-\frac{T}{NL}\right)^N=0,
\end{align}
a relationship inducing $\langle N_\emptyset\rangle\xrightarrow[]{T\to NL}0$: in words, the probability of observing empty hyperedges progressively vanishes as the density of $1$s increases. Consistently, in the sparse case one finds

\begin{align}
\lim_{T\to 0}p_\emptyset=\lim_{T\to 0}\left(1-\frac{T}{NL}\right)^N=1,
\end{align}
a relationship inducing $\langle N_\emptyset\rangle\xrightarrow[]{T\to0}L$: in words, the probability of observing empty hyperedges progressively rises as the density of $1$s decreases.\\

To evaluate the density of $1$s in the incidence matrix in correspondence of which the transition from the sparse to the dense regime happens, let us consider the case $N\gg 1$: more formally, this amounts to consider the asymptotic framework defined by letting $N\to+\infty$ while posing $T=O(L)$ \textcolor{black}{- equivalently, defined by posing} $p=O(1/N)$. Since $p=T/NL$ and $h=T/L$ remains finite, the probability for the generic hyperedge to be empty \textcolor{black}{obeys the relationship}

\begin{equation}
\textcolor{black}{\lim_{N\to+\infty}p_\emptyset=\lim_{N\to+\infty}\left(1-\frac{h}{N}\right)^N=e^{-h},}
\end{equation}
\textcolor{black}{i.e. remains finite as well: consistently, the expected number of empty hyperedges becomes}

\begin{equation}
\textcolor{black}{\lim_{N\to+\infty}\langle N_\emptyset\rangle=\lim_{N\to+\infty}Lp_\emptyset=Le^{-h};}
\end{equation}
upon imposing $Le^{-h}\leq1$, i.e. that the expected number of empty hyperedges is \emph{at most} 1, one derives what may be called \emph{filling threshold}, corresponding to $h_f^\text{RHM}\equiv\ln L$. In words, a value

\begin{equation}\label{eq:f_th_rhm}
\textcolor{black}{p>p_f^\text{RHM}=\frac{h_f^\text{RHM}}{N}=\frac{\ln L}{N}}
\end{equation}
ensures that the expected number of empty hyperedges in our random hypergraph is strictly less than one. \textcolor{black}{As a last observation, let us notice that evaluating $p_\emptyset$ in correspondence of the filling threshold returns the value $1/L$.}

\paragraph{A comparison with simple graphs: estimating the number of isolated nodes}

\textcolor{black}{A similar line of reasoning can be repeated for traditional graphs, the aim being, now, that of estimating $N_0$, i.e. the number of isolated nodes. To this aim, let us consider the asymptotic framework defined by letting $N\to+\infty$ while posing $L=O(N)$ \textcolor{black}{- equivalently, defined by posing} $q=O(1/N)$. Since $q=2L/N(N-1)$ and $k\equiv 2L/(N-1)$ remains finite, the probability for the generic node $i$ to be isolated obeys the relationship}

\begin{equation}
\textcolor{black}{\lim_{N\to+\infty}q_0=\lim_{N\to+\infty}(1-q)^{N-1}=\lim_{N\to+\infty}\left(1-\frac{k}{N-1}\right)^{N-1}=e^{-k},}
\end{equation}
i.e. remains finite as well: this, in turn, implies that the expected number of isolated nodes  $\langle N_0\rangle=Nq_0$ obeys the relationship

\begin{equation}
\textcolor{black}{\lim_{N\to+\infty}\langle N_0\rangle=\lim_{N\to+\infty}Nq_0=Ne^{-k};}
\end{equation}
\textcolor{black}{upon imposing $Ne^{-k}\leq1$, i.e. that the expected number of isolated nodes is \emph{at most} 1, one derives the \emph{connectivity threshold}, corresponding to $k_c^\text{RGM}\equiv\ln N$. In words, a value}

\begin{equation}
\textcolor{black}{q>q_c^\text{RGM}=\frac{k_c^\text{RGM}}{N}=\frac{\ln N}{N}}
\end{equation}
\textcolor{black}{ensures that the expected number of isolated nodes in our random graph is strictly less than one~\cite{Newman2010}. As a last observation, let us notice that evaluating $q_0$ in correspondence of the connectivity threshold returns the value $1/N$.}\\

\textcolor{black}{We also note that $N$ is the only quantity playing a relevant role in the case of graphs, while an interplay between $L$ and $N$ can be observed in the case of hypergraphs: in both cases, however, a condition on connectivity is present, driven by the request that the objects under investigations (nodes on the one hand and hyperedges on the other) have a non-zero number of connections.}

\paragraph{Estimation of the number of parallel hyperedges}

Let us, now, move to considering the issue of parallel hyperedges. By definition, two, parallel hyperedges $\alpha$ and $\beta$ are characterised by identical columns: hence, their Hamming distance, defined as the number of positions at which the corresponding symbols are different, is zero. More formally,

\begin{equation}
d_{\alpha\beta}\equiv\sum_{i=1}^N[I_{i\alpha}(1-I_{i\beta})+I_{i\beta}(1-I_{i\alpha})],
\end{equation}
a sum whose generic addendum is 1 in just two cases: either $I_{i\alpha}=1$ and $I_{i\beta}=0$ or $I_{i\alpha}=0$ and $I_{i\beta}=1$. Since $d_{\alpha\beta}\sim\text{Bin}(N,2p(1-p))$, one finds that 

\begin{equation}
p_{\parallel}^{\alpha\beta}\equiv P(d_{\alpha\beta}=0)=[1-2p(1-p)]^N
\end{equation}
and that

\begin{equation}
\langle d_{\alpha\beta}\rangle=2p(1-p)N
\end{equation}
$\forall\:\alpha\neq\beta$. Since $p(1-p)=(T/NL)(1-T/NL)$, one finds that

\begin{align}
\lim_{T\to NL}p(1-p)&=\lim_{T\to 0}p(1-p)=0,
\end{align}
i.e. $p(1-p)$ vanishes in both regimes, a result further implying that both $P(d_{\alpha\beta}=0)\xrightarrow[]{T\to NL}1$ and $P(d_{\alpha\beta}=0)\xrightarrow[]{T\to 0}1$ and that both $\langle d_{\alpha\beta}\rangle\xrightarrow[]{T\to NL}0$ and $\langle d_{\alpha\beta}\rangle\xrightarrow[]{T\to 0}0$: in words, the probability of observing parallel hyperedges progressively rises both as a consequence of having many $1$s and as a consequence of having few $1$s. Analogously for the expected Hamming distance.\\

Let us, now, evaluate the expected Hamming distance between any two hyperedges $\alpha$ and $\beta$ within the asymptotic framework defined by letting $N\to+\infty$ while posing $T=O(L)$ \textcolor{black}{- equivalently, defined by posing} $p=O(1/N)$. Since $p=T/NL$ and $h\equiv T/L$ remains finite, one finds that

\begin{equation}
\lim_{N\to+\infty}\langle d_{\alpha\beta}\rangle=\lim_{N\to+\infty}\frac{2h}{N}\left(1-\frac{h}{N}\right)N=2h;
\end{equation}
upon imposing $2h\geq1$, i.e. that the expected Hamming distance between any two hyperedges $\alpha$ and $\beta$ is \emph{at least} 1, one derives what may be called \emph{resolution threshold}, corresponding to $h_r^\text{RHM}\equiv 1/2$. In words, a value $p>p_r^\text{RHM}=h_r^\text{RHM}/N=1/2N$ ensures that, on average, any, two hyperedges $\alpha$ and $\beta$ differ by at least one element.\\

\textcolor{black}{To derive a  global condition on the total number of parallel hyperedges, note that, although the overlaps between pairs of hyperedges cannot be treated as i.i.d. random variables, the expected number of parallel hyperedges can still be computed explicitly. Upon posing $p_{\parallel}\equiv p_{\parallel}^{\alpha\beta}$, it reads}

\begin{equation}
\textcolor{black}{\langle N_{\parallel}\rangle=\sum_{\alpha=1}^L\sum_{\substack{\beta=1\\\beta>\alpha}}^Lp_{\parallel}=\frac{L(L-1)}{2}p_{\parallel}.}
\end{equation}
\textcolor{black}{ Considering that $p_{\parallel}\xrightarrow[]{N\to+\infty}e^{-2h}$ and imposing $\langle N_{\parallel}\rangle\leq1$, i.e. that the expected number of parallel hyperedges is \emph{at most} 1, one derives what may be called a \emph{multiple resolution threshold}, corresponding to $h_m^\text{RHM}\equiv\ln L-\ln\sqrt{2}\lesssim h_f^\text{RHM}$. In words, a value}

\begin{equation}
\textcolor{black}{p>p_m^\text{RHM}=\frac{h_m^\text{RHM}}{N}=\frac{\ln L}{N}-\frac{\ln\sqrt{2}}{N}\lesssim\frac{\ln L}{N}=\frac{h_f^\text{RHM}}{N}=p_f^\text{RHM}}
\end{equation}
\textcolor{black}{(also) ensures that the expected number of parallel hyperedges in our random hypergraph is strictly less than one.}

\paragraph{Estimation of the percolation threshold}

\textcolor{black}{The two thresholds derived in the previous subsections emerge in consequence of the attempts of solving the problems related to the appearance of empty as well as parallel hyperedges. Remarkably, a third threshold exists: known as \emph{percolation threshold}, it was, first, derived in~\cite{Barthelemy2022}, following the definition according to which any two hyperedges are said to be connected if they share at least one node. Here, we re-derive the percolation threshold by considering the `hypergraph to graph' projection (see also below): more formally, the total number of nodes shared by hyperedge $\alpha$ with any other hyperedge reads}

\begin{equation}
\textcolor{black}{\sigma_\alpha=\sum_{i=1}^N\sum_{\substack{\beta=1\\\beta\neq\alpha}}^LI_{i\alpha}I_{i\beta},}
\end{equation}
\textcolor{black}{its expected value being}

\begin{equation}
\textcolor{black}{\langle\sigma_\alpha\rangle=N(L-1)p^2\simeq NLp^2;}
\end{equation}
\textcolor{black}{imposing $\langle\sigma_\alpha\rangle=1$ leads to find the value}

\begin{equation}
\textcolor{black}{p_p^\text{RHM}=\frac{h_p^\text{RHM}}{N}=\frac{1}{\sqrt{NL}}}
\end{equation}
\textcolor{black}{which, in turn, induces the value $h_p^\text{RHM}\equiv\sqrt{N/L}$. In words, a value $p>p_p^\text{RHM}$ ensures that any two hyperedges in our random hypergraph share, on average, at least one node.}

\subsubsection*{Heterogeneous benchmarks:\\the Hypergraph Configuration Model (HCM)}

\paragraph{Microcanonical formulation}

The number of constraints can be enlarged to include the degrees, i.e. the sequence $\{k_i\}_{i=1}^N$, and the hyperdegrees, i.e. the sequence $\{h_\alpha\}_{\alpha=1}^L$. Although counting the number of configurations on which both sequences match their empirical values is a hard task, numerical recipes that shuffle the entries of a rectangular matrix, while preserving its marginals, exist~\cite{Strona2014,Carstens2015,Chodrow2020,kraakman2024}. \textcolor{black}{It should be, however, noticed that, if not carefully implemented, algorithms of the kind may lead to a non-uniform exploration of the space of configurations~\cite{Coolen2009,Roberts2012}; moreover, the issue concerning the time needed to collect a sufficiently large number of configurations should be addressed even in presence of an ergodic system. A recent proposal is that of extending the traditional Curveball algorithm to hypergraphs~\cite{kraakman2024}.}

\paragraph{Canonical formulation}

Solving the corresponding problem in the canonical framework is, instead, straightforward. Indeed, Shannon entropy maximisation leads to

\begin{align}\label{eq:p_HCM}
P(\mathbf{I})&=\frac{e^{-\sum_{i=1}^N\alpha_ik_i(\mathbf{I})-\sum_{\alpha=1}^L\beta_\alpha h_\alpha(\mathbf{I})}}{\sum_{\mathbf{I}\in\mathcal{I}}e^{-\sum_{i=1}^N\alpha_ik_i(\mathbf{I})-\sum_{\alpha=1}^L\beta_\alpha h_\alpha(\mathbf{I})}}=\frac{e^{-\sum_{i=1}^N\sum_{\alpha=1}^L(\alpha_i+\beta_\alpha)I_{i\alpha}}}{\sum_{\mathbf{I}\in\mathcal{I}}e^{-\sum_{i=1}^N\sum_{\alpha=1}^L(\alpha_i+\beta_\alpha)I_{i\alpha}}}\nonumber\\
&=\prod_{i=1}^N x_i^{k_i(\mathbf{I})}\prod_{\alpha=1}^L y_\alpha^{h_\alpha(\mathbf{I})}\prod_{i=1}^N\prod_{\alpha=1}^L(1+x_iy_\alpha)^{-1},
\end{align}
an expression that can be re-written as

\begin{align}
P(\mathbf{I})=\prod_{i=1}^N\prod_{\alpha=1}^Lp_{i\alpha}^{I_{i\alpha}}(1-p_{i\alpha})^{1-I_{i\alpha}}
\end{align}
with $e^{-\alpha_i}\equiv x_i$, $\forall\:i$, $e^{-\beta_\alpha}\equiv y_\alpha$, $\forall\:\alpha$ and $p_{i\alpha}\equiv x_iy_\alpha/(1+x_iy_\alpha)$, $\forall\:i,\alpha$. According to such a model, the entries of the incidence matrix of a hypergraph are independent random variables that obey different Bernoulli distributions, i.e. $I_{i\alpha}\sim\text{Ber}(p_{i\alpha})$, $\forall\:i,\alpha$. As a consequence, both degrees and hyperdegrees obey Poisson-Binomial distributions, i.e. $k_i\sim\text{PoissBin}(L,\{p_{i\alpha}\}_{\alpha=1}^L)$, $\forall\:i$ and $h_\alpha\sim\text{PoissBin}(N,\{p_{i\alpha}\}_{i=1}^N)$, $\forall\:\alpha$~\cite{Saracco2017}.

\paragraph{Parameter estimation}

In this case, solving the likelihood maximisation problem amounts to solve the system of coupled equations

\begin{align}
k_i^*&=\sum_{\alpha=1}^L\frac{x_iy_\alpha}{1+x_iy_\alpha}=\sum_{\alpha=1}^Lp_{i\alpha}=\langle k_i\rangle,\:\forall\:i\\
h_\alpha^*&=\sum_{i=1}^N\frac{x_iy_\alpha}{1+x_iy_\alpha}=\sum_{i=1}^Np_{i\alpha}=\langle h_\alpha\rangle,\:\forall\:\alpha
\end{align}
ensuring that $\langle k_i\rangle=k_i^*$, $\forall\:i$, $\langle h_\alpha\rangle=h_\alpha^*$, $\forall\:\alpha$ (and, as a consequence, $\langle T\rangle=T^*$). In case hypergraphs are sparse and in absence of hubs

\begin{equation}
p_{i\alpha}\simeq x_iy_\alpha=\frac{k_i^*h_\alpha^*}{T^*},\:\forall\:i,\alpha.
\end{equation}

The HCM reduces to a `partial' Configuration Model~\cite{Saracco2017} when either the degree or the hyperdegree sequence is left unconstrained (see also Section~\ref{app:hcm} in the Appendix). The canonical ensemble of each randomisation model (Table~\ref{tab2} in the Appendix sums up the sets of constraints defining them) can be explicitly sampled by considering each entry of $\mathbf{I}$, drawing a real number $u_{i\alpha}\in U[0,1]$ and posing $I_{i\alpha}=1$ if $u_{i\alpha}\leq p_{i\alpha}$, $\forall\:i,\alpha$.\\

The HCM is formally equivalent to the Bipartite Configuration Model (BiCM)~\cite{Saracco2015}. Such an identification is guaranteed by our focus on non-simple hypergraphs.

\paragraph{Estimation of the number of empty hyperedges}

Let us, now, consider the probability for the generic hyperedge $\alpha$ to be empty or, in other terms, that its hyperdegree $h_\alpha$ is zero. Upon remembering that $h_\alpha\sim\text{PoissBin}(N,\{p_{i\alpha}\}_{i=1}^N)$, one finds

\begin{equation}
p_\emptyset^\alpha\equiv\prod_{i=1}^N(1-p_{i\alpha})
\end{equation}
while the expected number of empty hyperedges, now, reads

\begin{equation}\label{eq.empty.hcm}
\langle N_\emptyset\rangle\equiv\sum_{\alpha=1}^Lp_\emptyset^\alpha=\sum_{\alpha=1}^L\prod_{i=1}^N(1-p_{i\alpha}).
\end{equation}

As previously done, let us inspect the behaviour of the aforementioned quantities on the ensemble induced by the HCM as the density of $1$s in the incidence matrix varies. Although it depends on (the heterogeneity of) the sets of coefficients $\{x_i\}_{i=1}^N$ and $\{y_\alpha\}_{\alpha=1}^L$, general conclusions can be still drawn within a simpler framework. To this aim, let us consider the functional form reading

\begin{equation}
p_{i\alpha}=\frac{zf_ig_\alpha}{1+zf_ig_\alpha},\:\forall\:i,\alpha
\end{equation}
where the vector of fitnesses $\{f_i\}_{i=1}^N$ accounts for the heterogeneity of nodes, the vector of fitnesses $\{g_\alpha\}_{\alpha=1}^L$ accounts for the heterogeneity of hyperedges and $z$ tunes the density of $1$s in the incidence matrix \textcolor{black}{- `partial' Configuration Models are recovered upon posing either $f_i=1$, $\forall\:i$ or $g_\alpha=1$, $\forall\:\alpha$.)} Within such a framework, the fitnesses of the nodes and the fitnesses of the hyperedges can be drawn from any distribution. The dense and sparse regimes are, now, defined by the positions $z\to+\infty$ and $z\to 0$, respectively. In the dense case, one finds

\begin{equation}
\lim_{z\to+\infty}p_\emptyset^\alpha=\lim_{z\to+\infty}\prod_{i=1}^N\left(1-\frac{zf_ig_\alpha}{1+zf_ig_\alpha}\right)=0,
\end{equation}
a relationship inducing $\langle N_\emptyset\rangle\xrightarrow[]{z\to+\infty}0$: in words, the probability of observing empty hyperedges progressively vanishes as the density of $1$s increases. Consistently, in the sparse case one finds

\begin{equation}
\lim_{z\to0}p_\emptyset^\alpha=\lim_{z\to0}\prod_{i=1}^N\left(1-\frac{zf_ig_\alpha}{1+zf_ig_\alpha}\right)=1,
\end{equation}
a relationship inducing $\langle N_\emptyset\rangle\xrightarrow[]{z\to0}L$: in words, the probability of observing empty hyperedges progressively rises as the density of $1$s decreases.\\

\textcolor{black}{For what concerns the filling threshold, a derivation that is similar-in-spirit to the one carried out for the case of the RHM can be sketched. Let us pose ourselves in the sparse regime: since $1-p_{i\alpha}\simeq e^{-p_{i\alpha}}$, the probability for the generic hyperedge to be empty satisfies the chain of relationships}

\begin{equation}
\textcolor{black}{p_\emptyset^\alpha=\prod_{i=1}^N(1-p_{i\alpha})\simeq\prod_{i=1}^N e^{-p_{i\alpha}}=e^{-\sum_{i=1}^Np_{i\alpha}}=e^{-h_\alpha};}
\end{equation}
\textcolor{black}{consistently, the expected number of empty hyperedges becomes}

\begin{equation}
\textcolor{black}{\langle N_\emptyset\rangle=\sum_{\alpha=1}^Lp_\emptyset^\alpha\simeq\sum_{\alpha=1}^L e^{-h_\alpha}}
\end{equation}
\textcolor{black}{and imposing $\langle N_\emptyset\rangle\leq1$, i.e. that the expected number of empty hyperedges is \emph{at most} 1, one derives a global condition to be satisfied by the hyperdegrees. In general terms, the aforementioned condition leads to require $e^{-h_\alpha}=O(1/L)$, i.e. $h_\alpha=O(\ln L)$, $\forall\:\alpha$ and $p_{i\alpha}=O(\ln L/N)$, $\forall\:i,\alpha$.}

\paragraph{A comparison with simple graphs: estimating the number of isolated nodes}

\textcolor{black}{Coming to traditional graphs, the aim is, now, to estimate the number of isolated nodes. In this case, $1-p_{ij}\simeq e^{-p_{ij}}$ and the probability for the generic node $i$ to be isolated reads}

\begin{equation}
\textcolor{black}{q_0^i=\prod_{\substack{j=1\\j\neq i}}^N(1-p_{ij})\simeq\prod_{\substack{j=1\\j\neq i}}^N e^{-p_{ij}}=e^{-\sum_{j(\neq i)=1}^Np_{ij}}=e^{-k_i};}
\end{equation}
\textcolor{black}{consistently, the expected number of isolated nodes becomes}

\begin{equation}
\textcolor{black}{\langle N_0\rangle=\sum_{i=1}^Nq_0^i\simeq\sum_{i=1}^N e^{-k_i}}
\end{equation}
\textcolor{black}{and imposing $\langle N_0\rangle\leq1$, i.e. that the expected number of isolated nodes is \emph{at most} 1, one derives a global condition to be satisfied by the degrees. In general terms, the aforementioned condition leads to require $e^{-k_i}=O(1/N)$, i.e. $k_i=O(\ln N)$, $\forall\:i$ and $p_{ij}=O(\ln N/N)$, $\forall\:i<j$.}

\paragraph{Estimation of the number of parallel hyperedges}

As in the case of the RHM, we consider the Hamming distance between the columns representing the two hyperedges $\alpha$ and $\beta$. Since, now, $d_{\alpha\beta}\sim\text{PoissBin}(N,\{q_i^{\alpha\beta}\}_{i=1}^N)$, where $q_i^{\alpha\beta}\equiv p_{i\alpha}(1-p_{i\beta})+p_{i\beta}(1-p_{i\alpha})$ with $p_{i\alpha}=zf_ig_\alpha/(1+zf_ig_\alpha)$ and $p_{i\beta}=zf_ig_\beta/(1+zf_ig_\beta)$, one finds that 

\begin{equation}
P(d_{\alpha\beta}=0)=\prod_{i=1}^N(1-q_i^{\alpha\beta})
\end{equation}
and that

\begin{equation}
\langle d_{\alpha\beta}\rangle=\sum_{i=1}^Nq_i^{\alpha\beta}
\end{equation}
$\forall\:\alpha\neq\beta$. Since

\begin{align}
\lim_{z\to+\infty}q_i^{\alpha\beta}&=\lim_{z\to0}q_i^{\alpha\beta}=0,
\end{align}
i.e. $q_i^{\alpha\beta}$ vanishes in both regimes, one finds that both $P(d_{\alpha\beta}=0)\xrightarrow[]{z\to+\infty}1$ and $P(d_{\alpha\beta}=0)\xrightarrow[]{z\to0}1$ and that both $\langle d_{\alpha\beta}\rangle\xrightarrow[]{z\to+\infty}0$ and $\langle d_{\alpha\beta}\rangle\xrightarrow[]{z\to0}0$: as in the case of the RHM, the probability of observing parallel hyperedges progressively rises both as a consequence of having many $1$s and as a consequence of having few $1$s. Analogously for the expected Hamming distance.\\

\textcolor{black}{For what concerns the resolution threshold, a derivation that is similar-in-spirit to the one carried out for the case of the RHM can be sketched. Let us pose ourselves in the sparse regime and consider that $1-q_i^{\alpha\beta}\simeq e^{-q_i^{\alpha\beta}}\simeq e^{-(p_{i\alpha}+p_{i\beta})}$. As a consequence}

\begin{equation}
\textcolor{black}{p_{\parallel}^{\alpha\beta}\equiv P(d_{\alpha\beta}=0)=\prod_{i=1}^N(1-q_i^{\alpha\beta})\simeq\prod_{i=1}^N e^{-q_i^{\alpha\beta}}\simeq e^{-\sum_{i=1}^N(p_{i\alpha}+p_{i\beta})}=e^{-(h_\alpha+h_\beta)}}
\end{equation}
\textcolor{black}{and}

\begin{equation}
\textcolor{black}{\langle d_{\alpha\beta}\rangle=\sum_{i=1}^Nq_i^{\alpha\beta}\simeq\sum_{i=1}^N(p_{i\alpha}+p_{i\beta})=h_\alpha+h_\beta.}
\end{equation}

\textcolor{black}{Imposing $\langle d_{\alpha\beta}\rangle\geq1$, i.e. that the expected Hamming distance between any two hyperedges $\alpha$ and $\beta$ is \emph{at least} 1, amounts to require that $P(d_{\alpha\beta}=0)\leq e^{-1}$ - thus recovering the same condition holding true in the case of the RHM where, in fact, $p_{\parallel}\equiv p_{\parallel}^{\alpha\beta}\xrightarrow[]{N\to+\infty}e^{-2h}$.}\\

\textcolor{black}{The expected number of parallel hyperedges, now, reads}

\begin{equation}
\textcolor{black}{\langle N_{\parallel}\rangle=\sum_{\alpha=1}^L\sum_{\substack{\beta=1\\\beta>\alpha}}^Lp_{\parallel}^{\alpha\beta}\simeq\sum_{\alpha=1}^L\sum_{\substack{\beta=1\\\beta>\alpha}}^Le^{-(h_\alpha+h_\beta)};}
\end{equation}
\textcolor{black}{upon imposing $\langle N_{\parallel}\rangle\leq1$, i.e. that the expected number of parallel hyperedges is \emph{at most} 1, one derives a global condition to be satisfied by the hyperdegrees. In general terms, the aforementioned condition leads to require $e^{-(h_\alpha+h_\beta)}=O(1/L^2)$, i.e. $h_\alpha=O(\ln L)$, $\forall\:\alpha$ and $p_{i\alpha}=O(\ln L/N)$, $\forall\:i,\alpha$.}

\paragraph{Estimation of the percolation threshold}

\textcolor{black}{For what concerns the percolation threshold, the expected value of the total number of nodes shared by hyperedge $\alpha$ with any other hyperedge, now, reads}

\begin{equation}
\textcolor{black}{\langle\sigma_\alpha\rangle=\sum_{i=1}^N\sum_{\substack{\beta=1\\\beta\neq\alpha}}^Lp_{i\alpha}p_{i\beta}}
\end{equation}
\textcolor{black}{and imposing $\langle\sigma_\alpha\rangle=1$ leads to a global condition to be satisfied. In general terms, the aforementioned condition leads to require $p_{i\alpha}p_{i\beta}=O(1/NL)$, i.e. $p_{i\alpha}=O(1/\sqrt{NL})$, $\forall\:i,\alpha$.}

\section*{Results}

\subsection*{Hypergraphs in the dense and sparse regime}

Let us start by verifying the correctness of the estimations of the \textcolor{black}{filling, multiple resolution and percolation thresholds} provided by our benchmarks: to this aim, we have considered the values $N=300$ and $L=1000$.

\subsubsection*{The Random Hypergraph Model}

\begin{figure*}[t!]
\includegraphics[width=\textwidth]{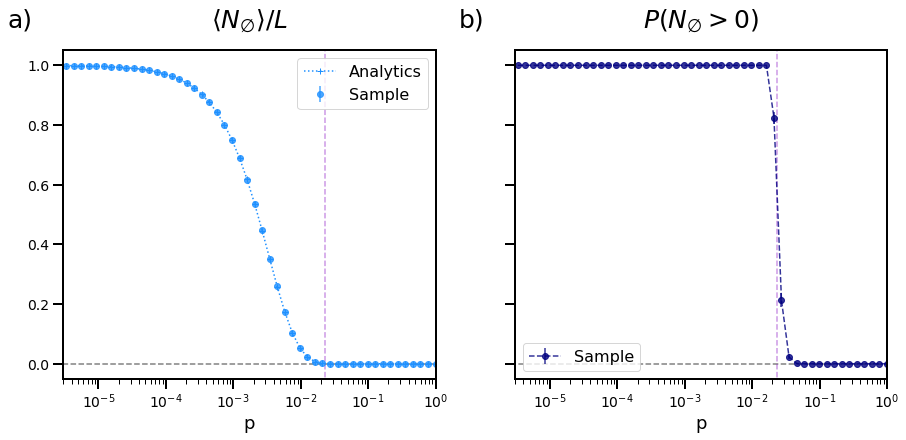}
\caption{\textcolor{black}{\textbf{Impact of empty hyperedges on the RHM ensemble}. More in detail, the t}rends of $\langle N_\emptyset\rangle/L=p_\emptyset=(1-p)^N\xrightarrow[]{N\to+\infty}e^{-h}$, i.e. the probability for the generic hyperedge to be emtpy \textcolor{black}{is represented in panel a} and the one of $P(N_\emptyset>0)=1-[1-(1-p)^N]^L\xrightarrow[]{N\to+\infty}1-(1-e^{-h})^L$, i.e. the probability of observing at least one empty hyperedge \textcolor{black}{is represented in panel b}. Evaluating them in correspondence of $p_f^\text{RHM}=h_f^\text{RHM}/N=\ln L/N\simeq 0.023$ (vertical line) returns, respectively, the values $1/L=10^{-3}$ and $1-(1-1/L)^L\simeq 0.6323$. The dense (sparse) regime is recovered for large (small) values of $p$. Each dot represents an average taken over an ensemble of $10^3$ configurations (explicitly sampled from the RHM) and is accompanied by the corresponding $95\%$ confidence interval.}
\label{fig2}
\end{figure*}

\begin{figure*}[t!]
\includegraphics[width=\textwidth]{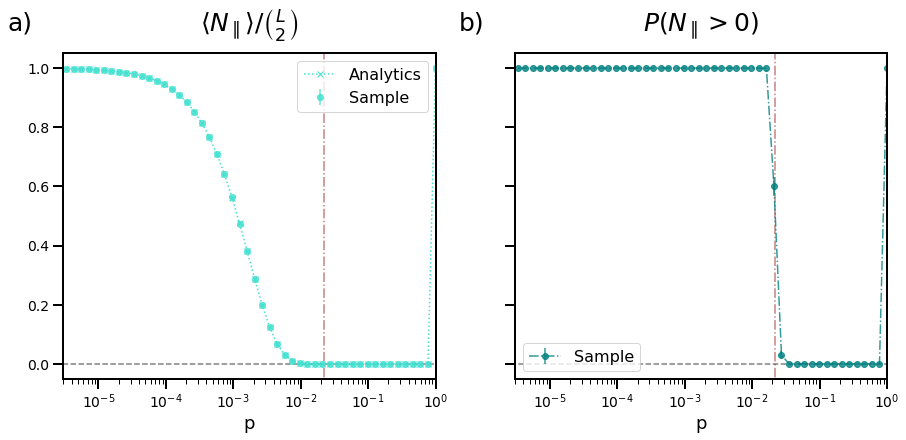}
\caption{\textcolor{black}{\textbf{Impact of parallel hyperedges on the RHM ensemble.} More in detail, the t}rends of $2\langle N_{\parallel}\rangle/L(L-1)=[1-2p(1-p)]^N=p_{\parallel}\xrightarrow[]{N\to+\infty}e^{-2h}$, i.e. the probability for the generic pair of hyperedges to be parallel and $P(N_{\parallel}>0)$, i.e. the probability of observing at least one pair of parallel hyperedges, as functions of $p$, \textcolor{black}{are represented, respectively, in panels a and b}. Evaluating them in correspondence of $p_m^\text{RHM}=h_m^\text{RHM}/N\lesssim\ln L/N\simeq 0.023$ (vertical line) returns, respectively, the values $1/L^2=10^{-6}$ and $\simeq 0.6$. The dense (sparse) regime is recovered for large (small) values of $p$. Each dot represents an average taken over an ensemble of $10^3$ configurations (explicitly sampled from the RHM) and is accompanied by the corresponding $95\%$ confidence interval.}
\label{fig3}
\end{figure*}

\begin{figure*}[t!]
\includegraphics[width=\textwidth]{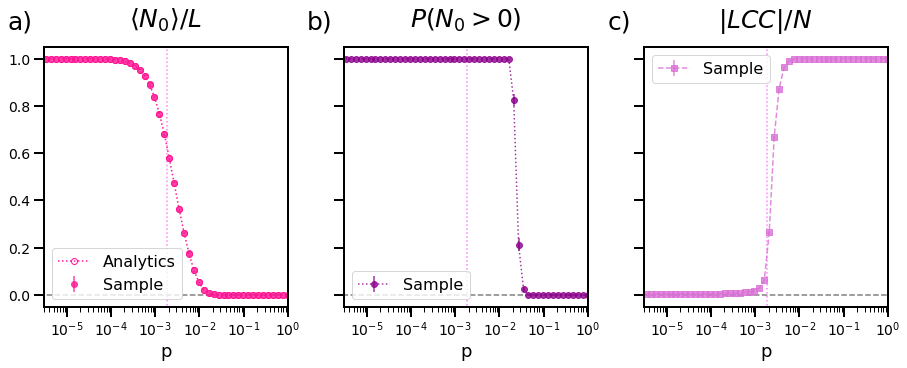}
\caption{\textcolor{black}{\textbf{Impact of isolated hyperedges on the RHM ensemble.} More in detail, t}rends of $\langle N_0\rangle/L=p_0$, i.e. the probability for the generic hyperedge to be isolated in the projection, $P(N_0>0)$, i.e. the probability of observing at least one, isolated hyperedge in the projection and $|\text{LCC}|/N$, i.e. the percentage of nodes belonging to the largest connected component (LCC), \textcolor{black}{are represented} as functions of $p$\textcolor{black}{, respectively in panels a, b and c}. Evaluating the first two in correspondence of $p_p^\text{RHM}=h_p^\text{RHM}/N=1/\sqrt{NL}\simeq 0.002$ (vertical line) returns, respectively, the values $e^{(e^{-s}-1)/s}\simeq 0.631$ and $\simeq 1$. The dense (sparse) regime is recovered for large (small) values of $p$. Each dot represents an average taken over an ensemble of $10^3$ configurations (explicitly sampled from the RHM) and is accompanied by the corresponding $95\%$ confidence interval.}
\label{fig4}
\end{figure*}

Each quantity has been plotted as a function of $p\in[10^{-6},1]$. The dense (sparse) regime is recovered for large (small) values of $p$. Each dot of Figs.~\ref{fig2},~\ref{fig3} and~\ref{fig4} represents an average taken over an ensemble of $10^3$ configurations explicitly sampled from the RHM and is accompanied by the corresponding $95\%$ confidence interval\textcolor{black}{, calculated via the bootstrap method~\cite{Efron1979}}.

\paragraph{The filling threshold}

\textcolor{black}{Panel a} of Fig.~\ref{fig2} depicts the (analytical) trend of $\langle N_\emptyset\rangle/L=p_\emptyset=(1-p)^N$ (solid line): its agreement with the numerical estimations (dots) confirms the correctness of our formula.

\textcolor{black}{Although the value of the filling threshold has been determined by inspecting the asymptotic behaviour of $\langle N_\emptyset\rangle$, the quantity showing the neatest transition from the sparse to the dense regime is the probability of observing at least one empty hyperedge}

\begin{equation}
\textcolor{black}{P(N_\emptyset>0)=1-P(N_\emptyset=0)=1-(1-p_\emptyset)^L=1-[1-(1-p)^N]^L},
\end{equation}
\textcolor{black}{where we have exploited the fact that $P(N_\emptyset>0)$ is nothing but the complementary of the probability that no hyperedge is empty. Since $p_\emptyset\xrightarrow[]{N\to+\infty}e^{-h}$, evaluating such an expression in correspondence of $h_f^\text{RHM}=\ln L\simeq 6.907$ returns the value $1/L=10^{-3}$ (see \textcolor{black}{panel a} of Fig.~\ref{fig2}). As a consequence, the result $P(N_\emptyset>0)\xrightarrow[]{N\to+\infty}1-(1-e^{-h})^L$ is numerically recovered (see \textcolor{black}{panel b} of Fig.~\ref{fig2}). In words, although the filling threshold ensures that each, single hyperedge is empty with an overall small probability, the likelihood of observing at least one, empty hyperedge is still large (i.e. $\simeq 2/3)$: the steepness of the trend of $P(N_\emptyset>0)$, however, suggests it to quickly vanish as the density of $1$s in the incidence matrix crosses the value $p_f^\text{RHM}=h_f^\text{RHM}/N=\ln L/N\simeq 0.023$.}

\paragraph{The multiple resolution threshold}

\textcolor{black}{Panel a} of Fig.~\ref{fig3} depicts the trend of $2\langle N_{\parallel}\rangle/L(L-1)=[1-2p(1-p)]^N=p_{\parallel}$ (solid line): again, its agreement with the numerical estimations (dots) confirms the correctness of our formula.

\textcolor{black}{Evaluating $p_{\parallel}\xrightarrow[]{N\to+\infty}e^{-2h}$ in correspondence of $h_m^\text{RHM}\lesssim\ln L\simeq 6.907$ returns the value $1/L^2=10^{-6}$ (see \textcolor{black}{panel a} of Fig.~\ref{fig3}): in words, the aforementioned, critical value causes the likelihood of observing any two parallel hyperedges to be almost the square of the probability for each single hyperedge to be empty (i.e. $1/L$, see the comments under Eq.~\ref{eq:f_th_rhm}).}

\textcolor{black}{As the overlaps between pairs of hyperedges cannot be treated as i.i.d. random variables, evaluating the probability of observing at least one pair of parallel hyperedges forces us to proceed in a purely numerical fashion. Calculating $P(N_{\parallel}>0)$ in correspondence of $p_m^\text{RHM}=h_m^\text{RHM}/N\lesssim\ln L/N\simeq 0.023$ returns the value $0.6$ (see the \textcolor{black}{panel b} of Fig.~\ref{fig3}).}\\

\textcolor{black}{Carrying out such an estimation in the aforementioned regime is, however, instructive as it leads to the expression $P(N_{\parallel}>0)=1-(1-p_{\parallel})^{L(L-1)/2}=1-\{1-[1-2p(1-p)]^N\}^{L(L-1)/2}\xrightarrow[]{N\to+\infty}1-[1-e^{-2h}]^{L(L-1)/2}$ that, evaluated in correspondence of $h_m^\text{RHM}\lesssim\ln L\simeq 6.907$, returns the value $P(N_{\parallel}>0)=1-(1-1/L^2)^{L(L-1)/2}\simeq 0.393$ - whose difference with $0.6$, obtained numerically, lets us fully appreciate the role played by correlations.} 

\paragraph{The percolation threshold}

\textcolor{black}{Let us, now, focus on the projection of our hypergraph onto the layer of hyperedges. The generic hyperedge is isolated either because is not `connected' with any node or because is a singleton (i.e. it is `connected' with a node with which no other hyperedge is `connected'): in symbols,}

\begin{equation}
\textcolor{black}{p_0\equiv \{1-p[1-(1-p)^{L-1}]\}^N=[(1-p)+p(1-p)^{L-1}]^N}.
\end{equation}

\textcolor{black}{In order to evaluate $p_0$ in correspondence of the percolation threshold, let us consider that $L=s^2N$, with $s^2=10/3$; one, then, finds}

\begin{equation}
\textcolor{black}{\lim_{N\to+\infty}p_0=\lim_{N\to+\infty}\left[\left(1-\frac{1}{Ns}\right)+\frac{1}{Ns}\left(1-\frac{1}{Ns}\right)^{s^2N-1}\right]^N=e^{-\frac{1-e^{-s}}{s}}},
\end{equation}
\textcolor{black}{whose numerical value amounts to $0.631$ (see \textcolor{black}{panel a} of Fig.~\ref{fig4}): in words, the value $p_0(p=p_p^\text{RHM})\lesssim 2/3$ implies that the expected number of isolated hyperedges in the projection $\langle N_0\rangle=Lp_0$ tends to $2L/3$ as $p$ tends to $p_p^\text{RHM}$.}

\textcolor{black}{Pairs of hyperedges cannot be treated as independent. Let us, in fact, consider the bipartite representation of a hypergraph: as adjacent pairs of hyperedges (say $\alpha$, $\beta$ and $\beta$, $\gamma$) may share some neighbours (on the opposite layer), the number of common neighbours of $\alpha$ and $\beta$ will, in general, covariate with the number of common neighbours of $\beta$ and $\gamma$; therefore, evaluating the probability of observing at least one, isolated hyperedge in the projection forces us to proceed in a purely numerical fashion.} Calculating $P(N_0>0)$ in correspondence of $p_p^\text{RHM}=h_p^\text{RHM}/N=1/\sqrt{NL}\simeq 0.002$ practically returns $1$ (see \textcolor{black}{panel b} of Fig.~\ref{fig4}). From the perspective of a hypergraph connectedness, the percolation threshold is `less strict' than the filling threshold, allowing for a larger number of disconnected nodes ($2/3$ of the total versus $1$).\\

\textcolor{black}{As before, estimating the percolation threshold in the regime where the pairs of hyperedges behave as i.i.d. Binomial random variables is instructive. In this case, projecting a bipartite networks onto the layer of hyperedges amounts to connect any two of them with probability $1-(1-p^2)^N$ - i.e. the complementary of the probability $(1-p^2)^N$ of not sharing any node. The number of isolated nodes, thus, obeys the relationship $N_0\sim\text{Bin}(L,p_0)$, with}

\begin{equation}
\textcolor{black}{p_0\equiv [(1-p^2)^N]^L}
\end{equation}
\textcolor{black}{being the probability for the generic hyperedge to be isolated: in words, such an expression returns the probability for the generic hyperedge to not share \emph{any} node - with probability $(1-p^2)^N$ - with \emph{any} other hyperedge - with a probability amounting to the previous one raised to the power of $L$. As a consequence, evaluating $p_0$ in correspondence of the percolation threshold returns a value tending to $1/3$, a result further implying that the expected number of isolated hyperedges in the projection $\langle N_0\rangle=Lp_0$ tends to the value $L/3$ - both letting us fully appreciate the role played by correlations.}

\textcolor{black}{Within such a context, the probability of observing at least one isolated hyperedge in the projection satisfies the chain of relationships $P(N_0>0)=1-P(N_0=0)=1-(1-p_0)^L=1-[1-(1-p^2)^{NL}]^L\xrightarrow[]{N\to+\infty}1-(1-e^{-1})^L$: as the last expression quickly converges to 1 for large values of $L$, the same qualitative behaviour observed before is thus recovered.}

Finally, one may wonder which kind of mesoscale structure is identified by the percolation threshold: the answer is provided by \textcolor{black}{panel c} of Fig.~\ref{fig4}, showing the appearance of a large connected component \textcolor{black}{- as also pointed out in~\cite{Barthelemy2022}, the presence of a large connected component is inspected by projecting our hypergraph onto the layer of \emph{nodes}, whose connectedness is ensured by requiring the connectedness of hyperedges.}

\paragraph{The role of thresholds in a hypergraph evolution}

\textcolor{black}{Let us, now, make a couple of observations. The first one concerns the result according to which $h_m^\text{RHM}\lesssim h_f^\text{RHM}$: such a relationship suggests that, while filling a large portion of the incidence matrix is not required to observe a limited amount of parallel hyperedges (see \textcolor{black}{panel b} of Fig.~\ref{fig3}), when empty hyperedges are no longer observed, parallel hyperedges are no longer observed as well.}

\textcolor{black}{The second one concerns the result according to which either $h_f^\text{RHM}\leq h_p^\text{RHM}$ or $h_f^\text{RHM}\geq h_p^\text{RHM}$. By progressively rising the parameter $p$, two, different thresholds are, thus, met: if $h_f^\text{RHM}\leq h_p^\text{RHM}$, the filling threshold $p_f^\text{RHM}=\ln L/N$ is met before the percolation threshold $p_p^\text{RHM}=1/\sqrt{NL}$, i.e. hyperedges are filled before they start sharing nodes - as a consequence, singletons appear; if $h_f^\text{RHM}\geq h_p^\text{RHM}$, the percolation threshold $p_p^\text{RHM}=1/\sqrt{NL}$ is met before the filling threshold $p_f^\text{RHM}=\ln L/N$, i.e. hyperedges start sharing nodes before they are filled - as a consequence, no singleton appears before the filling threshold is crossed.}

\textcolor{black}{Notice that, for simple graphs, $k_p^\text{RGM}=1\leq k_c^\text{RGM}=\ln N$, i.e. by progressively rising the parameter $q$, the percolation threshold is always met before the connectivity threshold.}

\subsubsection*{The Hypergraph Configuration Model}

\begin{figure*}[t!]
\includegraphics[width=\textwidth]{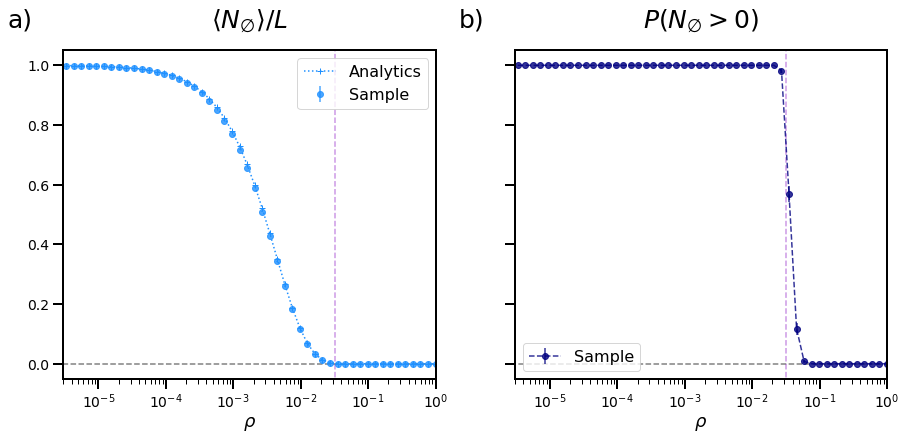}
\caption{\textcolor{black}{\textbf{Impact of empty hyperedges on the HCM ensemble}. More in detail, t}rends of $\langle N_\emptyset\rangle/L=\overline{p_\emptyset}$, i.e. the probability for the generic hyperedge to be empty and $P(N_\emptyset>0)\simeq 1-\prod_{\alpha=1}^L(1-e^{-h_\alpha})$, i.e. the probability of observing at least one empty hyperedge, \textcolor{black}{are represented} as functions of the connectance $\rho$\textcolor{black}{, respectively in panels a and b}. Evaluating the latter in correspondence of the filling threshold, reading $p_f^\text{HCM}\simeq 0.032$ (vertical line), returns the value $\simeq 0.642$. The dense (sparse) regime is recovered for large (small) values of $z$. Each dot represents an average taken over an ensemble of $10^3$ configurations (explicitly sampled from the HCM) and is accompanied by the corresponding $95\%$ confidence interval.}
\label{fig5}
\end{figure*}

\begin{figure*}[t!]
\includegraphics[width=\textwidth]{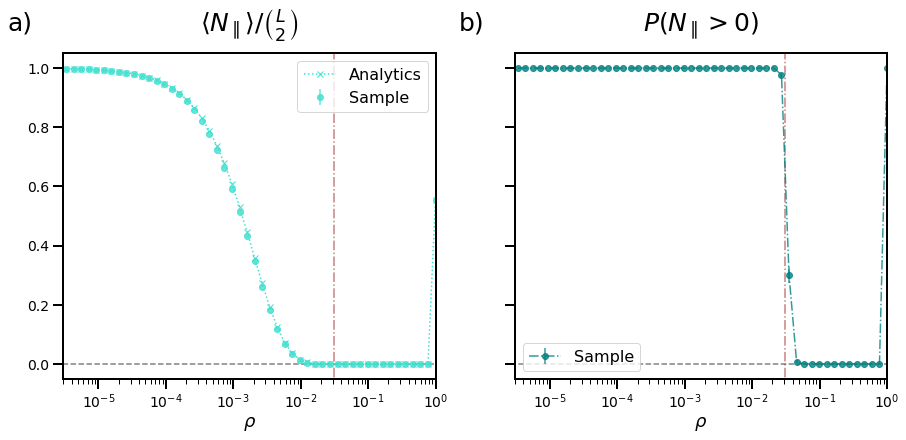}
\caption{\textcolor{black}{\textbf{Impact of parallel hyperedges on the HCM ensemble.} More in detail, t}rends of $2\langle N_{\parallel}\rangle/L(L-1)=\overline{p_{\parallel}}$, i.e. the probability for the generic pair of hyperedges to be parallel and $P(N_{\parallel}>0)$, i.e. the probability of observing at least one pair of parallel hyperedges, \textcolor{black}{are represented} as functions of the connectance $\rho$\textcolor{black}{, respectively in panels a and b}. Evaluating the latter in correspondence of the multiple resolution threshold, reading $p_m^\text{HCM}\simeq 0.031$ (vertical line), returns the $0.493$. The dense (sparse) regime is recovered for large (small) values of $z$. Each dot represents an average taken over an ensemble of $10^3$ configurations (explicitly sampled from the HCM) and is accompanied by the corresponding $95\%$ confidence interval.}
\label{fig6}
\end{figure*}

\begin{figure*}[t!]
\includegraphics[width=\textwidth]{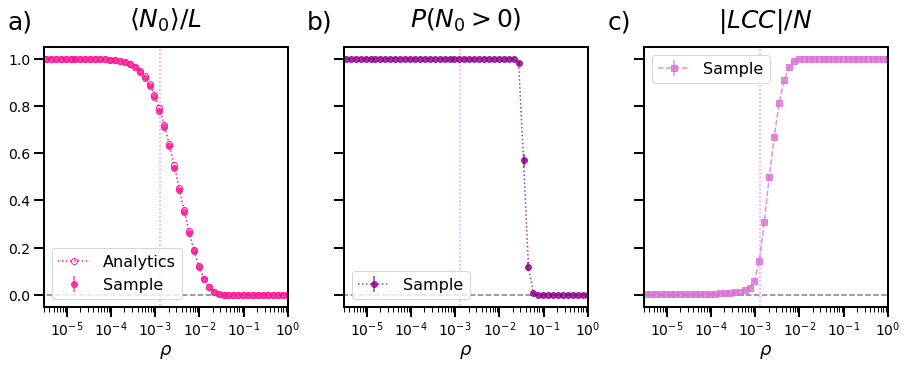}
\caption{\textcolor{black}{\textbf{Impact of isolated hyperedges on the HCM ensemble}. More in detail, t}rends of $\langle N_0\rangle/L=\overline{p_0}$, i.e. the probability for the generic hyperedge to be isolated in the projection, $P(N_0>0)$, i.e. the probability of observing at least one, isolated hyperedge in the projection, and $|\text{LCC}|/N$, i.e. the percentage of nodes belonging to the largest connected component (LCC), \textcolor{black}{are repesented} as functions of the connectance $\rho$\textcolor{black}{, respectively in panels a, b and c}. Evaluating the first two in correspondence of $p_p^\text{HCM}\simeq 0.001$ (vertical line), returns the values $\simeq 0.766$ and $\simeq 1$. The dense (sparse) regime is recovered for large (small) values of $p$. Each dot represents an average taken over an ensemble of $10^3$ configurations (explicitly sampled from the HCM) and is accompanied by the corresponding $95\%$ confidence interval.}
\label{fig7}
\end{figure*}

In order to carry out the numerical simulations in the case of the HCM, we have followed the procedure described in the previous sections and drawn both the fitnesses of nodes and those of hyperedges from a Pareto distribution with $\alpha=2$ \textcolor{black}{- other fat-tailed distributions were considered: qualitatively, results do not change.} Each quantity has been plotted as a function of $\rho(z)=\langle T\rangle/NL\in[10^{-6},1]$ where $\langle T\rangle=\sum_{i=1}^N\sum_{\alpha=1}^L zf_ig_\alpha/(1+zf_ig_\alpha)$ varies with $z$. The dense (sparse) regime is recovered for large (small) values of $z$. Each dot of Figs.~\ref{fig5},~\ref{fig6} and~\ref{fig7} represents an average taken over an ensemble of $10^3$ configurations explicitly sampled from the HCM and is accompanied by the corresponding $95\%$ confidence interval.

\paragraph{The filling threshold}

\textcolor{black}{Panel a} of Fig.~\ref{fig5} depicts the (analytical) trend of $\langle N_\emptyset\rangle/L=\sum_{\alpha=1}^Lp_\emptyset^\alpha/L=\overline{p_\emptyset}$ (solid line): as in the case of the RHM, its agreement with the numerical estimations (dots) confirms the correctness of our formula. Deriving an explicit expression for the filling threshold in the case of the HCM is a rather difficult task; still, we can proceed in a purely numerical fashion and individuate the value of the density of $1$s in the incidence matrix guaranteeing that the expected number of empty hyperedges (divided by $L$) amounts to $1$ (divided by $L$): although its, precise, numerical value depends on the values of the fitnesses, such a threshold still lies in the right tail of the trend induced by the HCM and reads $p_f^\text{HCM}\simeq 0.032$ (see \textcolor{black}{panel a} of Fig.~\ref{fig5}).

\textcolor{black}{Even in this case, the quantity showing the neatest transition from the sparse to the dense regime is the probability of observing at least one empty hyperedge}

\begin{equation}
\textcolor{black}{P(N_\emptyset>0)=1-P(N_\emptyset=0)=1-\prod_{\alpha=1}^L(1-p_\emptyset^\alpha)=1-\prod_{\alpha=1}^L\left[1-\prod_{i=1}^N(1-p_{i\alpha})\right]}
\end{equation}
\textcolor{black}{that, in the sparse regime, can be approximated as $P(N_\emptyset>0)\simeq 1-\prod_{\alpha=1}^L(1-e^{-h_\alpha})$: as \textcolor{black}{panel b} of Fig.~\ref{fig5} shows, evaluating $P(N_\emptyset>0)$ in correspondence of the filling threshold returns $0.642$. Finally, let us explicitly notice that the value of the filling threshold is shifted on the right with respect to its homogeneous counterpart, an evidence probably due to the presence of small fitnesses that increase the probability of observing at least one, empty hyperedge, hence requiring a larger value of $z$ to let $P(N_\emptyset>0)$ vanish.}

\paragraph{The multiple resolution threshold}

\textcolor{black}{\textcolor{black}{Panel a} of Fig.~\ref{fig6} depicts the (analytical) trend of $2\langle N_{\parallel}\rangle/L(L-1)=2\sum_{\alpha=1}^L\sum_{\substack{\beta=1\\\beta>\alpha}}^Lp_{\parallel}^{\alpha\beta}/L(L-1)=\overline{p_{\parallel}}$ (solid line): as in the case of the RHM, its agreement with the numerical estimations (dots) confirms the correctness of our formula.}

\textcolor{black}{Adopting the strategy described in the previous paragraph, i.e. that of requiring that the expected number of parallel hyperedges (divided by $L(L-1)/2$) amounts to $1$ (divided by $L(L-1)/2$), we found that $p_m^\text{HCM}\simeq 0.031$ (see \textcolor{black}{panel a} of Fig.~\ref{fig6}). Numerically clculating $P(N_{\parallel}>0)$ in correspondence of the multiple resolution threshold returns the value $0.493$ (see \textcolor{black}{panel b} of Fig.~\ref{fig6}).}

\textcolor{black}{Let us notice that the value of the multiple resolution threshold no longer coincides with the value of the filling threshold, although it is still shifted on the right with respect to its homogeneous counterpart.}

\paragraph{The percolation threshold}

\textcolor{black}{The probability for the generic hyperedge to be isolated in the projection, now, reads}

\begin{equation}
\textcolor{black}{p_0^\alpha\equiv \prod_{i=1}^N\left\{1-p_{i\alpha}\left[1-\prod_{\substack{\beta=1\\\beta\neq\alpha}}^L(1-p_{i\beta})\right]\right\}=\prod_{i=1}^N\left\{(1-p_{i\alpha})+p_{i\alpha}\prod_{\substack{\beta=1\\\beta\neq\alpha}}^L(1-p_{i\beta})\right\}}
\end{equation}
that, in the sparse regime, can be approximated as $p_0^\alpha\simeq\prod_{i=1}^N[(1-p_{i\alpha})+p_{i\alpha}e^{-k_i}]$. Adopting the strategy described in the previous paragraphs, i.e. that of requiring that the expected value of the total number of nodes shared by any hyperedge with any other hyperedge amounts to $1$ - in symbols, $\overline{\langle\sigma\rangle}=\sum_{\alpha=1}^L\langle\sigma_\alpha\rangle/L=1$ - we found that $p_p^\text{HCM}\simeq 0.001$: since $\langle N_0\rangle=\sum_{\alpha=1}^Lp_0^\alpha$, evaluating $\langle N_0\rangle/L=\sum_{\alpha=1}^Lp_0^\alpha/L=\overline{p_0}$ and $P(N_0>0)$ in correspondence of the percolation threshold returns respectively the values $0.766$ (see \textcolor{black}{panel a} of Fig.~\ref{fig7}) and $\simeq 1$ (see \textcolor{black}{panel b} of Fig.~\ref{fig7}). \textcolor{black}{For what concerns the hypergraph connectedness, the same conclusion drawn in the case of the RHM holds true, as the percolation threshold allows for a larger number of disconnected nodes ($3/4$ of the total versus $1$).}\\

\textcolor{black}{Analogously, the mesoscale structure individuated by the percolation threshold consists of a large connected component constituted (see \textcolor{black}{panel c} of Fig.~\ref{fig7}).}

\subsection*{Solving the HCM on real-world hypergraphs}

In order to test our benchmarks on real-world configurations, we have focused on a number of data sets taken from Austin R. Benson's website (\url{https://www.cs.cornell.edu/~arb/data/}), i.e. the \texttt{contact-primary-school}, the \texttt{email-Enron} and the \texttt{NDC-classes} ones.\\

\textcolor{black}{Although the parameters defining the HCM must be numerically determined by solving the system of equations induced by the likelihood maximisation, when the system under analysis is sparse they can be approximated as described in~\ref{app:hcm} of the Appendix. Such an approximation leads to the expression}

\begin{equation}\label{eq:31}
p_{i\alpha}\simeq x_iy_\alpha=\frac{k_i^*h_\alpha^*}{T^*},\:\forall\:i,\alpha
\end{equation}
\textcolor{black}{that, as Fig.~\ref{figA1} in the Appendix shows,} is quite accurate for each data set considered here - in fact, one can safely assume that $x_i\simeq k_i^*/\sqrt{T^*}$, $\forall\:i$ and $h_\alpha^*/\sqrt{T^*}$, $\forall\:\alpha$.

\subsection*{`Hypergraph to graph' projection}

The canonical formalism that we have adopted leads to factorisable distributions, i.e. distributions that can be written as a product of pair-wise probability distributions; this allows the expectation of several quantities of interest to be evaluated analytically.\\

\textcolor{black}{Let us start by considering the matrix, introduced in~\cite{Battiston2020}, reading}

\begin{equation}\label{eq:W}
\mathbf{W}=\mathbf{I}\cdot\mathbf{I}^T-\mathbf{K}
\end{equation}
with $\mathbf{K}$ being the diagonal matrix whose $i$-th entry reads $k_i$; according to the definition above, \textcolor{black}{it induces a projection of a hypergraph onto a weighted graph, whose generic entry}

\begin{figure}[t!]
\centering
\includegraphics[width=0.7\textwidth]{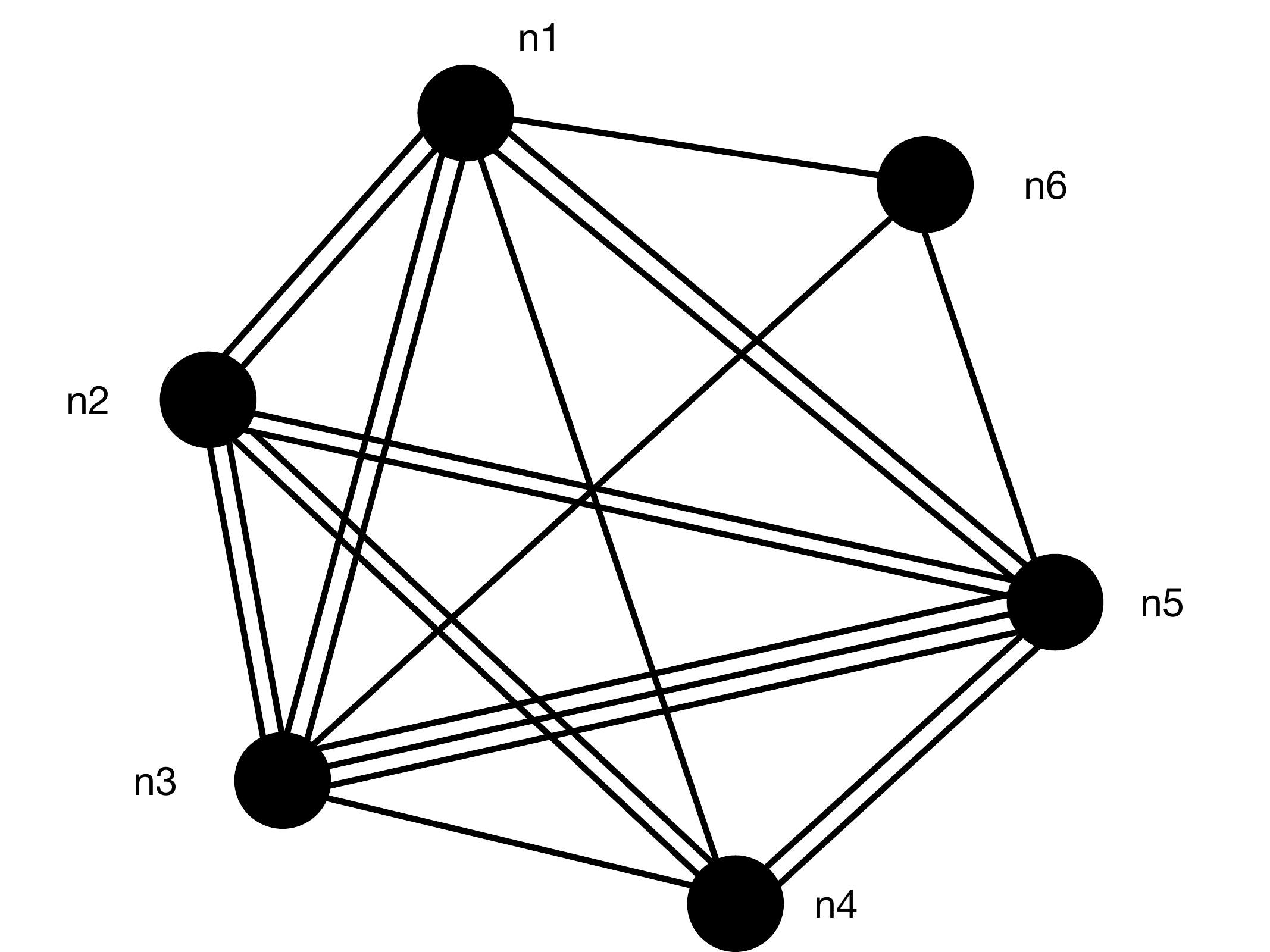}
\caption{\textbf{Graphical representation of the `hypergraph to graph' projection.} Notice that while the degree of each node (i.e. the number of hyperedges that are incident to it) reads $\{k_i\}=(3,3,3,3,4,1)$, the sigma of each node (i.e. the total number of hyperedges shared by it with any other node or, equivalently, the strength on the projection) reads $\{\sigma_i\}=(8,8,9,6,10,3)$ and the kappa of each node (i.e. the number of nodes it shares an hyperedge with or, equivalently, the degree on the projection) reads $\{\kappa_i\}=(5,4,5,4,5,3)$.}
\label{fig8}
\end{figure}

\begin{equation}
w_{ij}=\sum_{\alpha=1}^L I_{i\alpha}I_{j\alpha}-\delta_{ij}k_i
\end{equation}
\textcolor{black}{returns the number of hyperedges both $i$ and $j$ belong to} - more explicitly, $w_{ij}=\sum_{\alpha=1}^L I_{i\alpha}I_{j\alpha}$, $i\neq j$ and $w_{ii}=\sum_{\alpha=1}^L I_{i\alpha}I_{i\alpha}-k_i=\sum_{\alpha=1}^L I_{i\alpha}-k_i=k_i-k_i=0$. \textcolor{black}{In other words, $\mathbf{W}$ represents the object most closely resembling a traditional adjacency matrix.} The null models discussed so far can be employed to calculate $\langle w_{ij}\rangle$, $i\neq j$ that, in a perfectly general fashion, reads

\begin{equation}
\langle w_{ij}\rangle=\sum_{\alpha=1}^L\langle I_{i\alpha}I_{j\alpha}\rangle=\sum_{\alpha=1}^L\langle I_{i\alpha}\rangle\langle I_{j\alpha}\rangle=\sum_{\alpha=1}^Lp_{i\alpha}p_{j\alpha};
\end{equation}
\textcolor{black}{as we said,} the total number of hyperedges shared by node $i$ with any other node in the hypergraph (in a sense, its `strength' - see also Fig.~\ref{fig8}) can be computed as

\begin{equation}
\sigma_i=\sum_{\substack{j=1\\j\neq i}}^Nw_{ij}=\sum_{\substack{j=1\\j\neq i}}^N\sum_{\alpha=1}^LI_{i\alpha}I_{j\alpha}
\end{equation}
whose expected value reads

\begin{align}
\langle\sigma_i\rangle&=\sum_{\substack{j=1\\j\neq i}}^N\langle w_{ij}\rangle=\sum_{\alpha=1}^Lp_{i\alpha}[\langle h_\alpha\rangle-p_{i\alpha}].
\end{align}

As further confirmed by Fig.~\ref{figA1} in the Appendix, the approximation provided by Eq.~(\ref{eq:31}) allows us to pose $\langle\sigma_i\rangle_\text{HCM}\simeq k_i^*\sum_{\alpha=1}^L(h_\alpha^*)^2/T^*$, $\forall\:i$. Interestingly, as Fig.~\ref{fig9}a shows, the HCM overestimates the extent to which any two nodes of the \texttt{email-Enron} data set overlap: in words, such a real-world hypergraph is more compartmentalised than expected.\\

Let us, now, extend the concept of \emph{assortativity} to hypergraphs. To this aim, we consider the quantity named \emph{average incident hyperedges degree}, defined as

\begin{equation}
k_i^{nn}=\sum_{\alpha=1}^L\frac{I_{i\alpha}h_\alpha}{k_i}=\frac{\sigma_i+k_i}{k_i}=\frac{\sigma_i}{k_i}+1\simeq\frac{\sigma_i}{k_i}
\end{equation}
and representing the arithmetic mean of the degrees of the hyperedges including node $i$. An analytical approximation of its expected value can be provided as well:

\begin{align}
\langle k_i^{nn}\rangle&\simeq\sum_{\alpha=1}^L\frac{p_{i\alpha}[\langle h_\alpha\rangle+1-p_{i\alpha}]}{\langle k_i\rangle}=\frac{\langle\sigma_i\rangle+\langle k_i\rangle}{\langle k_i\rangle}=\frac{\langle\sigma_i\rangle}{\langle k_i\rangle}+1\simeq\frac{\langle\sigma_i\rangle}{\langle k_i\rangle}.
\end{align}

\subsection*{Disparity ratio and degree in the projection}

\begin{figure*}[t!]
\includegraphics[width=\textwidth]{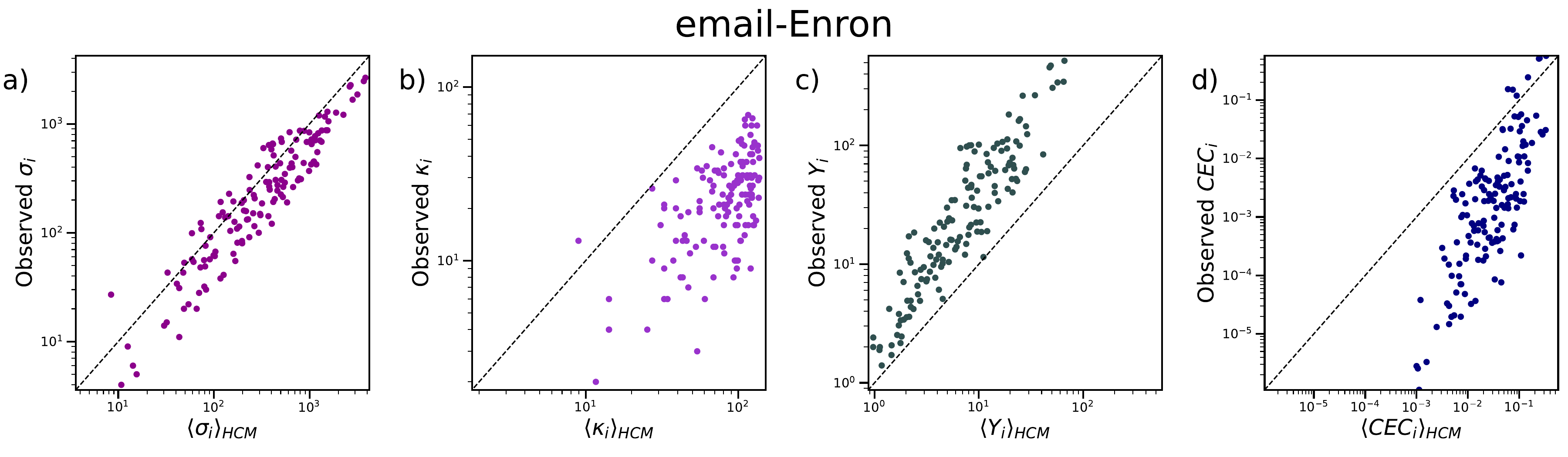}
\caption{\textcolor{black}{\textbf{Scatter plots between the empirical and the expected values for the \texttt{email-Enron} dataset.} More in detail, $\{\sigma_i\}$ vs. $\{\langle\sigma_i\rangle_\text{HCM}\}$ (panel a), $\{\kappa_i\}$ vs. $\{\langle\kappa_i\rangle_\text{HCM}\}$ (panel b), $\{Y_i\}$ vs. $\langle\{Y_i\rangle_\text{HCM}\}$ (panel c), and $\{\text{CEC}_i\}$ vs. $\{\langle\text{CEC}_i\rangle_\text{HCM}\}$ (panel d).} The HCM overestimates the extent to which any, two nodes overlap, as well as the CEC; the disparity ratio, instead, is underestimated by it. These results can be understood by considering that the HCM just constrains the degree sequences, hence inducing an ensemble where connections are `distributed' more evenly than observed.}
\label{fig9}
\end{figure*}

More information about the patterns shaping real-world hypergraphs can be obtained upon defining the ratio $f_{ij}=w_{ij}/\sigma_i$, $i\neq j$ that induces the quantity

\begin{equation}\label{eq:mortacci}
Y_i=\sum_{\substack{j=1\\j\neq i}}^Nf_{ij}^2=\sum_{\substack{j=1\\j\neq i}}^N\frac{w_{ij}^2}{\sigma_i^2},
\end{equation}
known as \emph{disparity ratio} and quantifying the (un)evenness of the distribution of the weights constituting the strength of node $i$ over the $\kappa_i=\sum_{\substack{j=1\\(j\neq i)}}^N\Theta[w_{ij}]\equiv\sum_{\substack{j=1\\j\neq i}}^Na_{ij}$ links characterising its connectivity - since $a_{ij}=1$ if nodes $i$ and $j$ share, at least, one hyperedge, $\kappa_i$ is the degree of node $i$ in the projection of the hypergraph (see also Fig.~\ref{fig8} and Fig.~\ref{fig9}b). Since, under the RHM, $w_{ij}\sim\text{Bin}(L,p^2)$, we find that $\langle a_{ij}\rangle=1-(1-p^2)^L$, i.e. the expected value of $a_{ij}$ coincides with the probability of observing a non-zero overlap. Under the HCM, instead, $w_{ij}\sim\text{PoissBin}(L,\{p_{i\alpha}p_{j\alpha}\}_{\alpha=1}^L)$, hence

\begin{equation}\label{eq_temp}
\langle a_{ij}\rangle=1-\prod_{\alpha=1}^L(1-p_{i\alpha}p_{j\alpha}).
\end{equation}

Let us also notice that

\begin{equation}
Y_i=\frac{1}{\kappa_i}
\end{equation}
in case weights are equally distributed among the connections established by node $i$, i.e. $w_{ij}=a_{ij}\sigma_i/\kappa_i$, $i\neq j$. Any larger value signals an excess concentration of weight in one or more links. An analytical approximation of the expected value of the disparity ratio of node $i$ can be provided as well:

\begin{equation}
\langle Y_i\rangle\simeq\sum_{\substack{j=1\\j\neq i}}^N\frac{\langle w_{ij}^2\rangle}{\langle\sigma_i^2\rangle}.
\end{equation}

Contrarily to what has been previously observed, the expected value of the disparity ratio cannot be always safely decomposed as a ratio of expected values - not even if the `full' HCM is employed. In fact, while this approximation works relatively well for the \texttt{contact-primary-school} data set, it does not for the \texttt{email-Enron} and the \texttt{NDC-classes} ones (see also Fig.~\ref{figB3} in the Appendix). For this reason, the expected value of the disparity ratio has been evaluated by explicitly sampling the ensemble of incidence matrices induced by the `full' HCM. In any case, as Fig.~\ref{fig9}c shows, such a null model underestimates the disparity ratio characterising each node of the \texttt{email-Enron} data set: in words, the empirical overlap between any two nodes is (much) less evenly `distributed' than expected. A similar conclusion can be drawn by considering $\langle\kappa_{i}\rangle=\sum_{i=1}^N\left[1-\prod_{\alpha=1}^L(1-p_{i\alpha}p_{j\alpha})\right]$: as Fig.~\ref{fig9}b shows, the degree of the nodes in the projection tend to be significantly smaller than expected, meaning that hyperedges concentrate on fewer edges than expected. This observation is in line with recent works showing the encapsulation and `simpliciality' of real-world hypergraphs~\cite{larock2023encapsulation,landry2024simpliciality}.

\subsection*{Eigenvector centrality}

Centrality measures for hypergraphs have been defined as well. An example is provided by the \emph{clique motif eigenvector centrality} (CEC), defined in~\cite{Benson2019} (see also Section~\ref{app:cec} of the Appendix): $\text{CEC}_i$ corresponds to the $i$-th entry of the Perron-Frobenius eigenvector of $\mathbf{W}$. As Fig.~\ref{fig9}d shows, the HCM underestimates the CEC as well: such a result can be understood by considering that the HCM constrains only the degree sequences, hence inducing an ensemble where connections are `distributed' more evenly than observed, an evidence letting the nodes overlap more, thus causing the entries of $\langle\mathbf{W}\rangle$ to be overall larger and less dissimilar, as well as those of its Perron-Frobenius eigenvector.

\subsection*{Confusion matrix}

Let us, now, consider the set of indices constituting the so-called \emph{confusion matrix} (see also Section~\ref{app:conf} of the Appendix). \textcolor{black}{They are intended to quantify the capability of a given network model in reproducing microscopic properties such as the position of $1$s and $0$s by explicitly comparing their empirical location with the one expected under the chosen model.} They are named \emph{true positive rate} (TPR), i.e. the percentage of $1$s correctly recovered by a given method, whose expected value reads

\begin{equation}
\langle\text{TPR}\rangle=\sum_{i=1}^N\sum_{\alpha=1}^L\frac{I_{i\alpha}p_{i\alpha}}{T};
\end{equation}
\emph{specificity} (SPC), i.e. the percentage of 0s correctly recovered by a given method, whose expected value reads

\begin{equation}
\langle\text{SPC}\rangle=\sum_{i=1}^N\sum_{\alpha=1}^L\frac{(1-I_{i\alpha})(1-p_{i\alpha})}{NL-T};
\end{equation}
\emph{positive predictive value} (PPV), i.e. the percentage of $1$s correctly recovered by a given method with respect to the total number of $1$s predicted by it, whose expected value reads

\begin{equation}
\langle\text{PPV}\rangle=\sum_{i=1}^N\sum_{\alpha=1}^L\frac{I_{i\alpha}p_{i\alpha}}{\langle T\rangle};
\end{equation}
\emph{accuracy} (ACC), measuring the overall performance of a given method in correctly placing both $1$s and $0$s, whose expected value reads

\begin{equation}
\langle\text{ACC}\rangle=\frac{\langle\text{TP}\rangle+\langle\text{TN}\rangle}{NL}.
\end{equation}

Results on the confusion matrix of a number of real-world hypergraphs reveal that the large sparsity of the latter ones makes it difficult to reproduce the TPR and the PPV (see also Table~\ref{tab:confusion} in the Appendix); on the other hand, the capability of the HCM (both in its `full' and approximated version) to reproduce the density of $1$s - and, as a consequence, the density of $0$s - ensures the SPC to be recovered quite precisely, in turn ensuring the overall accuracy of the model to be large (for an overall evaluation of the performance of the HCM in reproducing real-world hypergraphs, see also Table~\ref{tab:R2} in Section~\ref{app:r2} of the Appendix).

\begin{figure*}[t!]
\includegraphics[width=\textwidth]{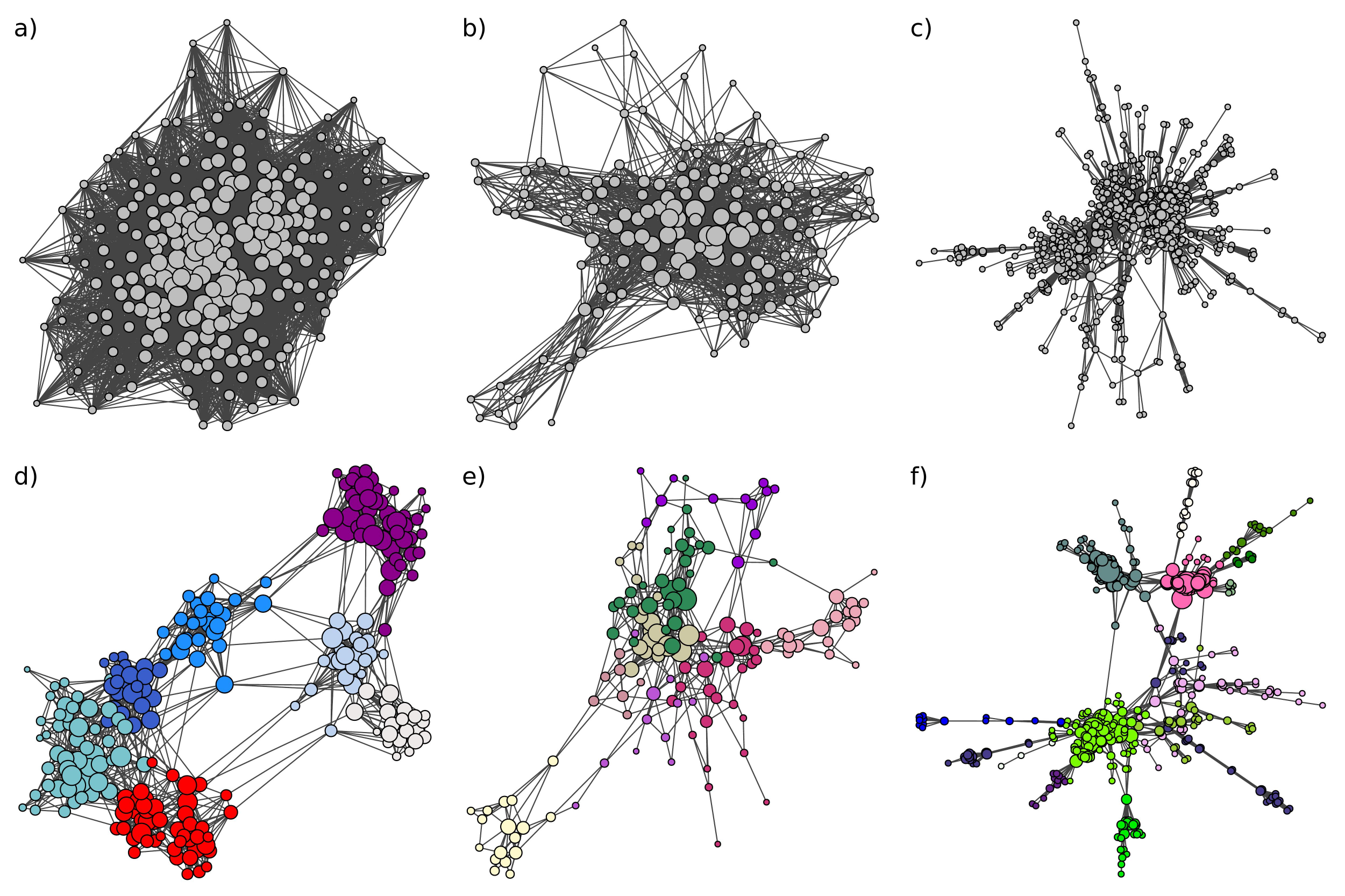}
\caption{\textcolor{black}{\textbf{Validated vs. non-validated 'hypergraph to network' projections of empirical datasets.} Panels a, b and c: projections of the \texttt{contact-primary-school}, \texttt{email-Enron} and \texttt{NDC-classes} data sets onto the layer of nodes. Panels d, e and f: validated counterparts of the aforementioned projections: any two nodes are linked if they share a significantly large number of hyperedges.} Communities have been detected by running the Louvain algorithm.}
\label{fig10}
\end{figure*}

\subsection*{Community detection}

Communities are commonly understood as densely connected groups of nodes. Representing an hypergraph via its incidence matrix allows this statement to be made more precise from a statistical perspective: in fact, the null models discussed so far can be employed to test if any two nodes share a significantly large number of hyperedges - hence can be clustered together, should this be the case. \textcolor{black}{In other words, it is possible to devise a `validation procedure' that filters the projection described by the matrix $\mathbf{W}$ by removing the entries that do not satisfy the requirement above.}

To this aim, we can adapt the recipe proposed in~\cite{Saracco2017} to project bipartite networks and summarised in the following. One, first, computes

\begin{equation}
\text{p-value}(w_{ij}^*)=\sum_{x\geq w_{ij}^*}f(x)
\end{equation}
for each pair of nodes; $f(x)$ depends on the chosen null model: in case the RHM is employed, it coincides with the Binomial distribution $\text{Bin}(x|L,p)$; in case the HCM is employed, it coincides with the Poisson-Binomial distribution $\text{PoissBin}(x|L,\{p_{i\alpha}p_{j\alpha}\}_{\alpha=1}^L)$. Second, one implements the FDR procedure, designed to handle multiple tests of hypothesis~\cite{Benjamini1995}: in practice, after ranking the p-values in increasing order, i.e. $\text{p-value}_1\leq\text{p-value}_2\leq\dots\leq\text{p-value}_n$, one individuates the largest integer $\hat{i}$ satisfying the condition
\begin{equation}
\text{p-value}_{\hat{i}}\leq\frac{\hat{i}t}{n}
\end{equation}
where $n=N(N-1)/2$ and $t$ is the single-test significance level, set to $0.01$ in the present analysis. Third, one links the (pairs of) nodes whose related p-value is smaller than the aforementioned threshold.\\

Fig.~\ref{fig10} shows the partitions returned by the Louvain algorithm run on the validated projections: as noticed elsewhere~\cite{Saracco2017,Pratelli2024,Guarino2024}, the detection of mesoscale structures is enhanced if carried out on filtered topologies.

\section*{Conclusions}

Our paper contributes to current research on hypergraphs by extending the constrained entropy-maximisation framework to incidence matrices, i.e. their simplest, tabular representation. Differently from the currently-available techniques~\cite{Chodrow2020}, our methodology has the advantage of being analytically tractable, scalable and versatile enough to be straightforwardly extensible to directed and/or weighted hypergraphs.

Beside leading to results whose relevance is mostly theoretical (i.e. the individuation of different regimes for higher-order structures and the estimation of the actual impact of empty and parallel hyperedges on the analysis of empirical systems), our models prove to be particularly useful when employed as benchmarks for real-world systems, i.e. for detecting patterns that are not imputable to purely random effects. Specifically, our results suggest that real-world hypergraphs are characterised by a degree of self-organisation that is absolutely non-trivial (see also Section~\ref{app:r2} of the Appendix).

This is even more surprising when considering that our results are obtained under a benchmark such as the HCM, i.e. a null model constraining both the degree and the hyperdegree sequences: since it overestimates the extent to which any two nodes overlap - a result whose relevance becomes evident as soon as one considers the effects that higher-order structures have on spreading and cooperation processes~\cite{de2021multistability,iacopini2022group,st2022influential} - our future efforts will be directed towards the analysis of benchmarks constraining non-linear quantities such as the co-occurrences between nodes and/or hyperedges.

\bmhead{Acknowledgements}

RL acknowledges support from the EPSRC grants EP/V013068/1, EP/V03474X/1 and EP/Y028872/1. TS acknowledges support from SoBigData.it that receives funding from European Union – NextGenerationEU – National Recovery and Resilience Plan (Piano Nazionale di Ripresa e Resilienza, PNRR) – Project: `SoBigData.it – Strengthening the Italian RI for Social Mining and Big Data Analytics' - Prot. IR0000013 – Avviso n. 3264 del 28/12/2021. FS acknowledges support from the project `CODE - Coupling Opinion Dynamics with Epidemics', funded under PNRR Mission 4 `Education and Research' - Component C2 - Investment 1.1 - Next Generation EU `Fund for National Research Program and Projects of Significant National Interest' PRIN 2022 PNRR, grant code P2022AKRZ9.

\section*{Declarations}

%\subsection*{Funding}

%RL acknowledges support from the EPSRC grants EP/V013068/1, EP/V03474X/1 and EP/Y028872/1. TS acknowledges support from SoBigData.it that receives funding from European Union – NextGenerationEU – National Recovery and Resilience Plan (Piano Nazionale di Ripresa e Resilienza, PNRR) – Project: `SoBigData.it – Strengthening the Italian RI for Social Mining and Big Data Analytics' - Prot. IR0000013 – Avviso n. 3264 del 28/12/2021. FS acknowledges support from the project `CODE - Coupling Opinion Dynamics with Epidemics', funded under PNRR Mission 4 `Education and Research' - Component C2 - Investment 1.1 - Next Generation EU `Fund for National Research Program and Projects of Significant National Interest' PRIN 2022 PNRR, grant code P2022AKRZ9.

\subsection*{Competing interests}

The authors declare no competing interests.

%\subsection*{Ethics approval and consent to participate}

%Not applicable

%\subsection*{Consent for publication}

%All authors approved the present version of the manuscript.

\subsection*{Data availability}

All the data used for the analysis are freely available at \url{https://www.cs.cornell.edu/~arb/data/}.

%\subsection*{Materials availability}

%Not applicable.

\subsection*{Code availability}

The authors will provide the code used for the analysis upon request.

\subsection*{Author contribution}

F.S., G.P., R.L. and T.S. conceived the study. F.S. performed the analyses and T.S. wrote the first version of the manuscript. F.S., G.P., R.L. and T.S. discussed the results, revised the draft and approved the final version of the manuscript.

\appendix 
\section{The Hypergraph Configuration Model}\label{app:hcm}

\subsection{Approximating the HCM}

\begin{figure}[t!]
\includegraphics[width=\textwidth]{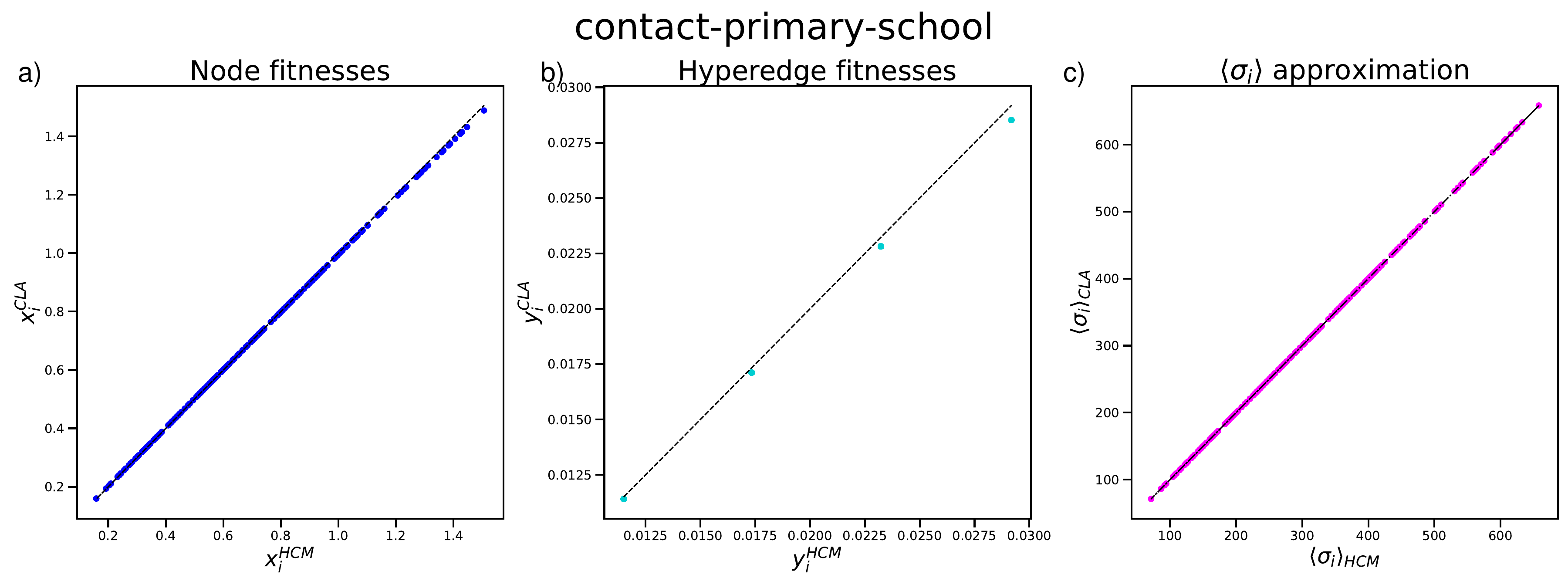}
\includegraphics[width=\textwidth]{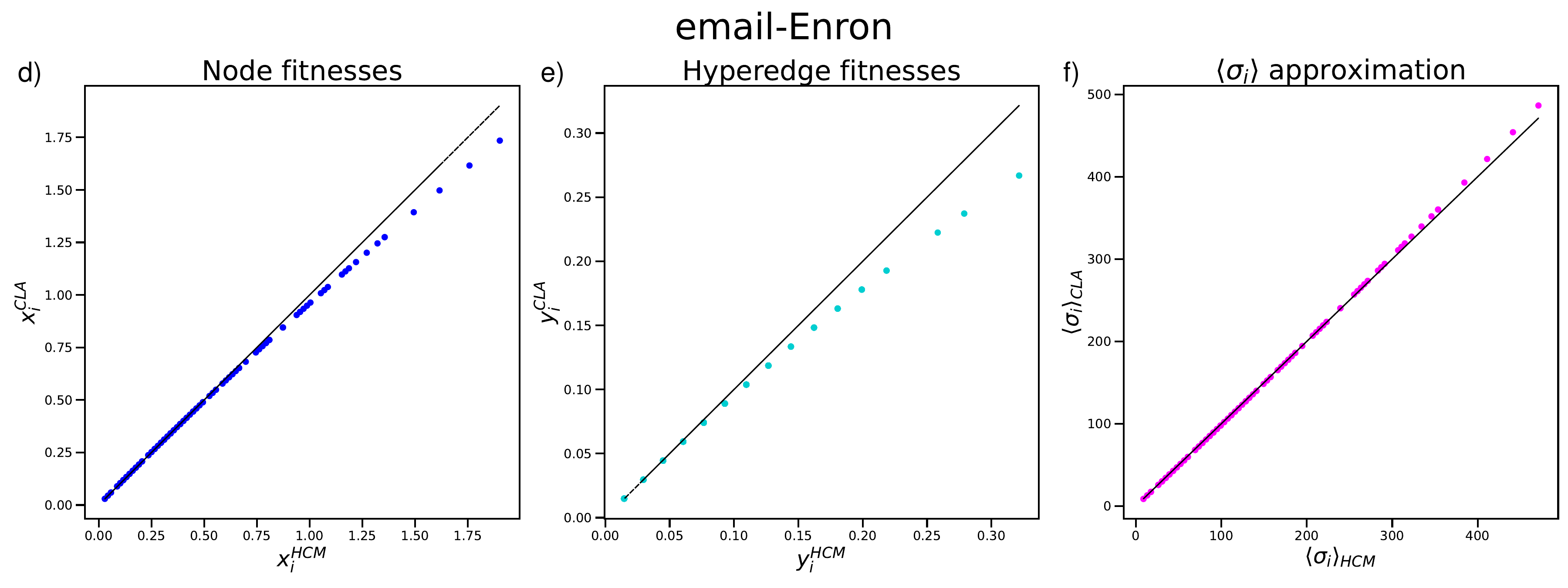}
\includegraphics[width=\textwidth]{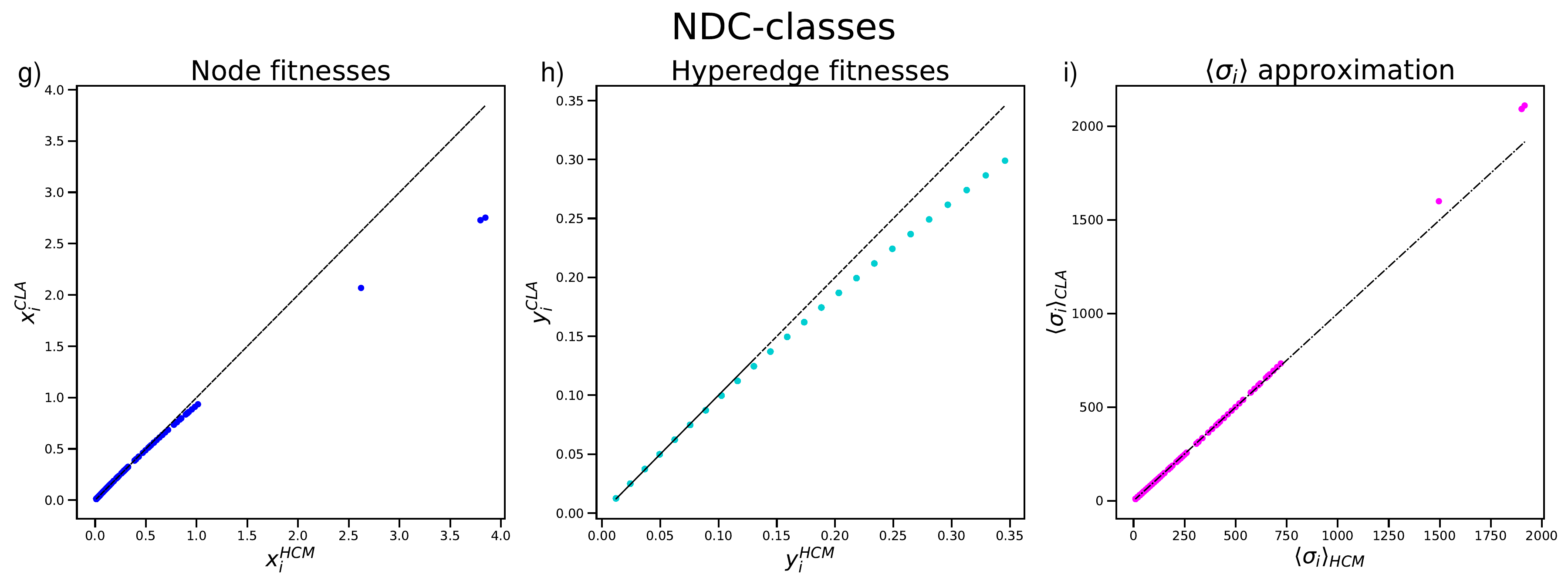}
\caption{\textbf{Effectiveness of Chung-Lu approximation.} Panels a, d and g: numerical values of the node-specific Lagrange multipliers obtained by solving the `full' HCM scattered versus the ones derived by adopting the CLA. Panels b, e and h: numerical values of the hyperedge-specific Lagrange multipliers obtained by solving the `full' HCM scattered versus the ones derived by adopting the CLA. Panels c, f and i: expected values $\{\langle\sigma_i\rangle_\text{HCM}\}$ scattered versus the expected values $\{\langle\sigma_i\rangle_\text{CLA}\}$.}
\label{figA1}
\end{figure}

Upon considering that the generic probability coefficient induced by the HCM can be Taylor-expanded as

\begin{equation}
p_{i\alpha}^\text{HCM}=\frac{x_iy_\alpha}{1+x_iy_\alpha}=x_iy_\alpha-(x_iy_\alpha)^2+(x_iy_\alpha)^3\dots
\end{equation}
one may wonder at which order the expansion can be safely truncated. In case hypergraphs are sparse, i.e. $T/NL\ll1$, and in absence of hubs, it turns out that just considering the first order is enough to obtain predictions as accurate as those achievable under the `full' HCM: in other words, one can simply put $p_{i\alpha}^\text{FOA}\simeq x_iy_\alpha$. Upon doing so, the system of equations defining the HCM simplifies to

\begin{align}
k_i^*&=\sum_{\alpha=1}^Lx_iy_\alpha,\:\forall\:i\\
h_\alpha^*&=\sum_{i=1}^Nx_iy_\alpha,\:\forall\:\alpha
\end{align}
expressions leading us to find $x_i=k_i^*/\sqrt{T^*}$, $\forall\:i$ and $y_\alpha=h_\alpha^*/\sqrt{T^*}$, $\forall\:\alpha$. As a consequence, $p_{i\alpha}^\text{FOA}=x_iy_\alpha=k_i^*h_\alpha^*/T^*$ - a position that is commonly named Chung-Lu approximation (CLA). Fig.~\ref{figA1} shows the goodness of the latter in approximating $\langle\sigma_i\rangle_\text{HCM}$ for the \texttt{contact-primary-school}, the \texttt{email-Enron} and the \texttt{congress-bills} data sets, i.e. the accuracy of the following chain of equalities

\begin{align}
\langle\sigma_i\rangle_\text{HCM}&=\sum_{\alpha=1}^Lp_{i\alpha}^\text{HCM}[h_\alpha-p_{i\alpha}^\text{HCM}]\nonumber\\
&\overset{\text{FOA}}{\simeq}\sum_{\alpha=1}^Lx_iy_\alpha[h_\alpha-x_iy_\alpha]\nonumber\\
&\overset{\text{CLA}}{\simeq}\frac{k_i^*}{T^*}\left(1-\frac{k_i^*}{T^*}\right)\sum_{\alpha=1}^L(h_{\alpha}^*)^2\nonumber\\
&\simeq k_i^*\sum_{\alpha=1}^L\frac{(h_{\alpha}^*)^2}{T^*},\:\forall\:i.
\end{align}

\clearpage

\begin{table}[t!]
\captionsetup{width=\textwidth}
\caption{\textbf{Summary of the constraints defining the canonical version of each benchmark considered here.} While the total number of $1$s is preserved by each model, the degrees are separately preserved by the two, partial HCMs and jointly preserved only by the HCM. Stars indicate the empirical values of the constraints.}
\label{tab2}
\centering
\begin{tabular}{l|c|c|c}
\hline
& $\langle T\rangle$ & $\langle k_i\rangle$ & $\langle h_\alpha\rangle$ \\
\hline
\hline
Random Hypergraph Model & $T^*$ & $T^*/N$ & $T^*/L$ \\
Partial Configuration Model - nodes layer & $T^*$ & $k_i^*$ & $T^*/L$ \\
Partial Configuration Model - hyperedges layer & $T^*$ & $T^*/N$ & $h_\alpha^*$ \\
Hypergraph Configuration Model & $T^*$ & $k_i^*$ & $h_\alpha^*$ \\
\hline
\end{tabular}
\end{table}

\subsection{Partial Configuration Models}

The HCM reduces to a `partial' Configuration Model in case one of the two degree sequences is left unconstrained.\\

\textit{Partial Configuration Model - nodes layer.} Let us start by solely constraining the degree sequence $\{k_i\}_{i=1}^N$, by posing $\beta_\alpha=0$, $\forall\:\alpha$ (or, equivalently, $y_\alpha=1$, $\forall\:\alpha$). Our canonical probability distribution becomes

\begin{equation}
P(\mathbf{I})=\prod_{i=1}^N\prod_{\alpha=1}^L\frac{x_i^{I_{i\alpha}}}{1+x_i}=\prod_{i=1}^Np_i^{k_i}(1-p_i)^{L-k_i}
\end{equation}
where $e^{-\alpha_i}\equiv x_i$ and $p_i\equiv x_i/(1+x_i)$, $\forall\:i$. The entries of the incidence matrix are, now, independent random variables obeying the Bernoulli distributions reading

\begin{equation}
I_{i\alpha}\sim\text{Ber}(p_i),\:\forall\:i,\alpha;
\end{equation}
in words, the entries along the same row obey the same distribution while the ones along the same column obey different distributions. As a consequence, the $i$-th node degree (a sum of i.i.d. Bernoulli random variables), obeys the Binomial distribution

\begin{equation}
k_i\sim\text{Bin}(L,p_i)
\end{equation}
while the $\alpha$-th hyperedge degree (a sum of independent random variables that obey different Bernoulli distributions) obeys the Poisson-Binomial distribution 

\begin{equation}
h_\alpha\sim\text{PoissBin}(N,p_1\dots p_N).
\end{equation}

Interestingly, while the degrees of the nodes obey different Binomial distributions, the degrees of the hyperedges obey the same Poisson-Binomial distribution.\\

The resolution of the likelihood maximisation problem leads us to find the values $p_i=k_i^*/L$, $\forall\:i$ which, in turn, ensure that $\langle k_i\rangle_{\text{HPCM}_\text{nodes}}=\sum_{\alpha=1}^Lp_i=k_i^*$, $\forall\:i$ and that $\langle T\rangle_{\text{HPCM}_\text{nodes}}=T^*$; instead, $\langle h_\alpha\rangle_{\text{HPCM}_\text{nodes}}$ will, in general, differ from $h_\alpha^*$ - in fact, $\langle h_\alpha\rangle_{\text{HPCM}_{\text{nodes}}}=T^*/L$, $\forall\:\alpha$.\\

From a microcanonical perspective, the number of configurations satisfying the requirement that the degrees of the nodes match their empirical values amounts to

\begin{equation}
\Omega_{\text{HPCM}_\text{nodes}}=\prod_{i=1}^N\binom{L}{k_i^*};
\end{equation}
reshuffling the $1$s along each row of the incidence matrix separately ensures the degrees of the nodes to be preserved while destroying any other correlation.\\

\textit{Partial Configuration Model - hyperedges layer.} Analogously, the canonical probability distribution describing the case in which only the hyperdegree sequence $\{h_\alpha\}_{\alpha=1}^L$ is constrained reads

\begin{equation}
P(\mathbf{I})=\prod_{i=1}^N\prod_{\alpha=1}^L\frac{y_\alpha^{I_{i\alpha}}}{1+y_\alpha}=\prod_{\alpha=1}^Lp_\alpha^{h_\alpha}(1-p_\alpha)^{N-h_\alpha}
\end{equation}
where $e^{-\beta_\alpha}\equiv y_\alpha$ and $p_\alpha\equiv y_\alpha/(1+y_\alpha)$, $\forall\:\alpha$. As for the previous null model, the entries of the incidence matrix are independent random variables obeying different Bernoulli distributions, i.e.

\begin{equation}
I_{i\alpha}\sim\text{Ber}(p_\alpha),\:\forall\:i,\alpha;
\end{equation}
in words, the entries along the same column obey the same distribution while the ones along the same row obey different distributions. As a consequence, the $i$-th node degree (a sum of independent random variables that obey different Bernoulli distributions) obeys the Poisson-Binomial distribution 

\begin{equation}
k_i\sim\text{PoissBin}(L,p_1\dots p_L)
\end{equation}
while the $\alpha$-th hyperedge degree (a sum of i.i.d. Bernoulli random variables) obeys the Binomial distribution 

\begin{equation}
h_\alpha\sim\text{Bin}(N,p_\alpha).
\end{equation}

Interestingly, while the degrees of the nodes obey the same Poisson-Binomial distribution, the degrees of the hyperedges obey different Binomial distributions.\\

The resolution of the likelihood maximisation problem leads us to find the values $p_\alpha=h_\alpha^*/N$, $\forall\:\alpha$ which, in turn, ensure that $\langle h_\alpha\rangle_{\text{HPCM}_\text{hyperedges}}=\sum_{i=1}^Np_\alpha=h_\alpha^*$, $\forall\:\alpha$ and that $\langle T\rangle_{\text{HPCM}_\text{hyperedges}}=T^*$; instead, $\langle k_i\rangle_{\text{HPCM}_\text{hyperedges}}$ will, in general, differ from $k_i^*$ - in fact, $\langle k_i\rangle_{\text{HPCM}_\text{hyperedges}}=\sum_{\alpha=1}^Lp_\alpha=T^*/N$, $\forall\:i$.\\

From a microcanonical perspective, the number of configurations satisfying the requirement that the degrees of the hyperedges match their empirical values amounts to

\begin{equation}
\Omega_{\text{HPCM}_\text{hyperedges}}=\prod_{\alpha=1}^L\binom{N}{h_\alpha^*};
\end{equation}
reshuffling the $1$s along each column of the incidence matrix separately ensures the degrees of the hyperedges to be preserved, while destroying any other correlation.

\begin{figure*}[t!]
\includegraphics[width=\textwidth]{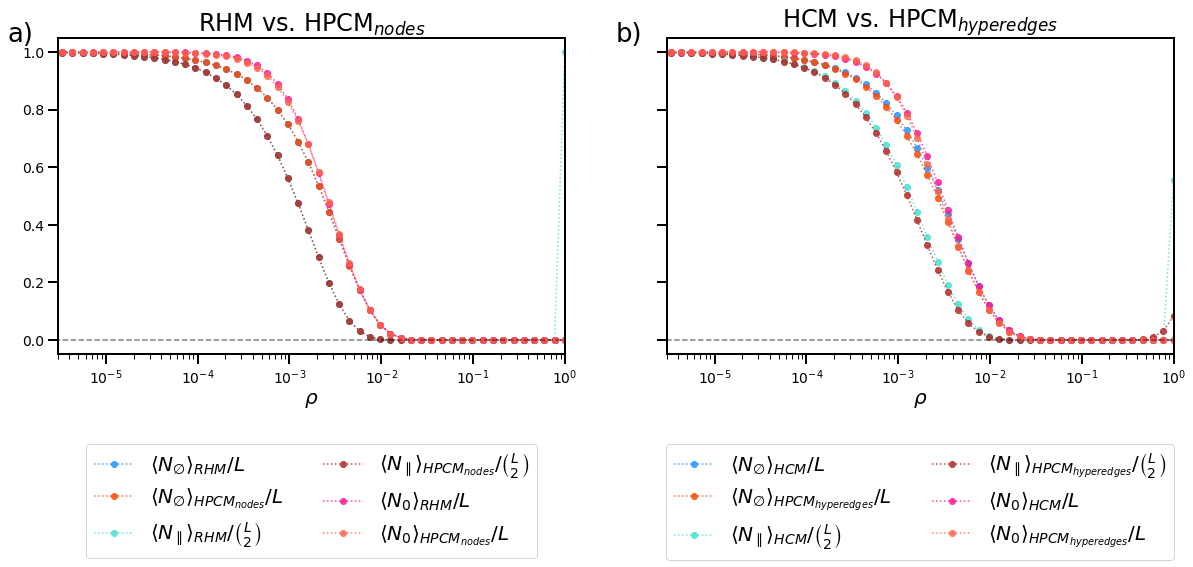}
\caption{\textbf{Comparison of the impact of empty, parallel or isolated hyperedges in RHM, HCM and partial models.} More in detail, trends of $\langle N_\emptyset\rangle/L$, i.e. the probability for the generic hyperedge to be empty, $2\langle N_{\parallel}\rangle/L(L-1)$, i.e. the probability for the generic pair of hyperedges to be parallel and $\langle N_0\rangle/L$, i.e. the probability for the generic hyperedge to be isolated in the projection, as functions of $\rho$, under the `partial' Configuration Models: while the trends returned by HPCM$_\text{nodes}$ overlap with those of the RHM (panel a), the trends returned by HPCM$_\text{hyperedges}$ overlap with those of the `full' HCM (panel b). The dense (sparse) regime is recovered for large (small) values of $p$. The vertical lines indicate the density values corresponding to the filling threshold and to the multiple resolution threshold.}
\label{figA2}
\end{figure*}

\subsection{Asymptotic results}

As noticed in the main text, the asymptotic behaviour of the HCM depends on the whole set of coefficients $\{p_{i\alpha}\}_{i,\alpha}$: closed form expressions can be, thus, obtained only after having specified the functional dependence of each of them on $N$. General conclusions can be drawn more easily for the two, `partial' Configuration Models.

Let us start by considering the behaviour of HPCM$_\text{hyperedges}$, defined by the coefficients $p_{i\alpha}=h_\alpha/N$, $\forall\:i,\alpha$. Such a position leads to the expressions

\begin{align}
p_\emptyset^\alpha&=\prod_{i=1}^N(1-p_{i\alpha})=\left(1-\frac{h_\alpha}{N}\right)^N\rightarrow e^{-h_\alpha},\\
p_{\parallel}^{\alpha\beta}&=\prod_{i=1}^N(1-q_i^{\alpha\beta})\simeq\left(1-\frac{h_\alpha+h_\beta}{N}\right)^N\rightarrow e^{-(h_\alpha+h_\beta)},
\end{align}
\textcolor{black}{proving that the estimations concerning the number of empty as well as parallel hyperedges returned by HPCM$_\text{hyperedges}$ are compatible with those returned by the `full' HCM and}

\begin{align}
p_0^\alpha=\prod_{i=1}^N\left\{1-p_{i\alpha}+p_{i\alpha}\prod_{\substack{\beta=1\\\beta\neq\alpha}}^L(1-p_{i\beta})\right\}\simeq \left[1-\frac{h_\alpha}{N}(1-e^{-T/N})\right]^N\rightarrow e^{-h_\alpha(1-e^{-T/N})},
\end{align}
\textcolor{black}{whose compatibility with the prediction returned by the `full' HCM has been inspected only numerically (see Fig.~\ref{figA2}).}

Let us, now, inspect the behaviour of HPCM$_\text{nodes}$, defined by the coefficients $p_{i\alpha}=k_i/L$, $\forall\:i,\alpha$. Such a position, in the sparse regime, leads to the expressions

\begin{align}
p_\emptyset^\alpha&=\prod_{i=1}^N(1-p_{i\alpha})=\prod_{i=1}^N\left(1-\frac{k_i}{L}\right)\simeq \prod_{i=1}^Ne^{-\frac{k_i}{L}}=e^{-\frac{\sum_{i=1}^Nk_i}{L}}=e^{-\frac{T}{L}}=e^{-h},\\
p_{\parallel}^{\alpha\beta}&=\prod_{i=1}^N(1-q_i^{\alpha\beta})=\prod_{i=1}^N\left(1-\frac{2k_i}{L}\right)\simeq \prod_{i=1}^Ne^{-\frac{2k_i}{L}}=e^{-\frac{2\sum_{i=1}^Nk_i}{L}}=e^{-\frac{2T}{L}}=e^{-2h},
\end{align}
\textcolor{black}{proving that the estimations concerning the number of empty as well as parallel hyperedges returned by HPCM$_\text{nodes}$ are compatible with those returned by the RHM and}

\begin{align}
p_0^\alpha=\prod_{i=1}^N\left\{1-p_{i\alpha}+p_{i\alpha}\prod_{\substack{\beta=1\\\beta\neq\alpha}}^L(1-p_{i\beta})\right\}\simeq \prod_{i=1}^N\left[1-\frac{k_i}{L}\left(1-e^{-k_i}\right)\right]\simeq e^{-\frac{\sum_{i=1}^Nk_i(1-e^{-k_i})}{L}},
\end{align}
\textcolor{black}{whose compatibility with the prediction returned by the RHM has been inspected only numerically (see Fig.~\ref{figA2}).}

\clearpage

\section{Expected value of topological quantities}

Let us, now, provide the explicit expression of the expected value of some topological properties of interest. Let us start with the entries of the matrix $\mathbf{W}$, for which the following results hold true

\begin{align}
\langle w_{ij}\rangle&=\sum_{\alpha=1}^L\langle I_{i\alpha}I_{j\alpha}\rangle=\sum_{\alpha=1}^L\langle I_{i\alpha}\rangle\langle I_{j\alpha}\rangle=\sum_{\alpha=1}^Lp_{i\alpha}p_{j\alpha},\\
\text{Var}[w_{ij}]&=\sum_{\alpha=1}^L\text{Var}[I_{i\alpha}I_{j\alpha}]+2\sum_{\beta=1}^L\sum_{\substack{\gamma=1\\\gamma>\beta}}^L\text{Cov}[I_{i\beta}I_{j\beta},I_{i\gamma}I_{j\gamma}]=\sum_{\alpha=1}^Lp_{i\alpha}p_{j\alpha}(1-p_{i\alpha}p_{j\alpha}).
\end{align}

Therefore,

\begin{align}
\sum_{\substack{j=1\\j\neq i}}^N\langle w_{ij}\rangle^2&=\sum_{\alpha=1}^Lp_{i\alpha}^2\left[\sum_{j=1}^Np_{j\alpha}^2-p_{i\alpha}^2\right]+2\sum_{\beta=1}^L\sum_{\substack{\gamma=1\\\gamma>\beta}}^Lp_{i\beta}p_{i\gamma}\left[\sum_{j=1}^Np_{j\beta}p_{j\gamma}-p_{i\beta}p_{i\gamma}\right],\\
\sum_{\substack{j=1\\j\neq i}}^N\text{Var}[w_{ij}]&=\sum_{\alpha=1}^Lp_{i\alpha}(\langle h_\alpha\rangle-p_{i\alpha})-\sum_{\alpha=1}^Lp_{i\alpha}^2\left[\sum_{j=1}^Np_{j\alpha}^2-p_{i\alpha}^2\right].
\end{align}

\begin{figure}[t!]
\includegraphics[width=\textwidth]{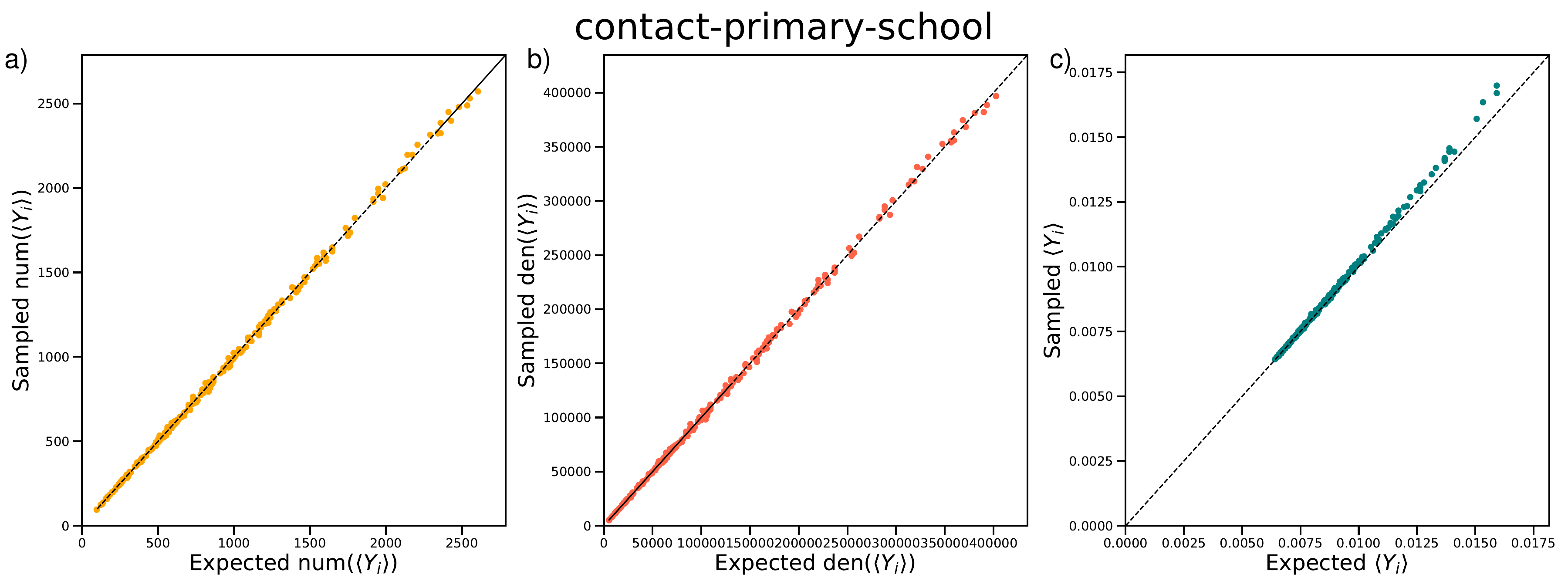}
\includegraphics[width=\textwidth]{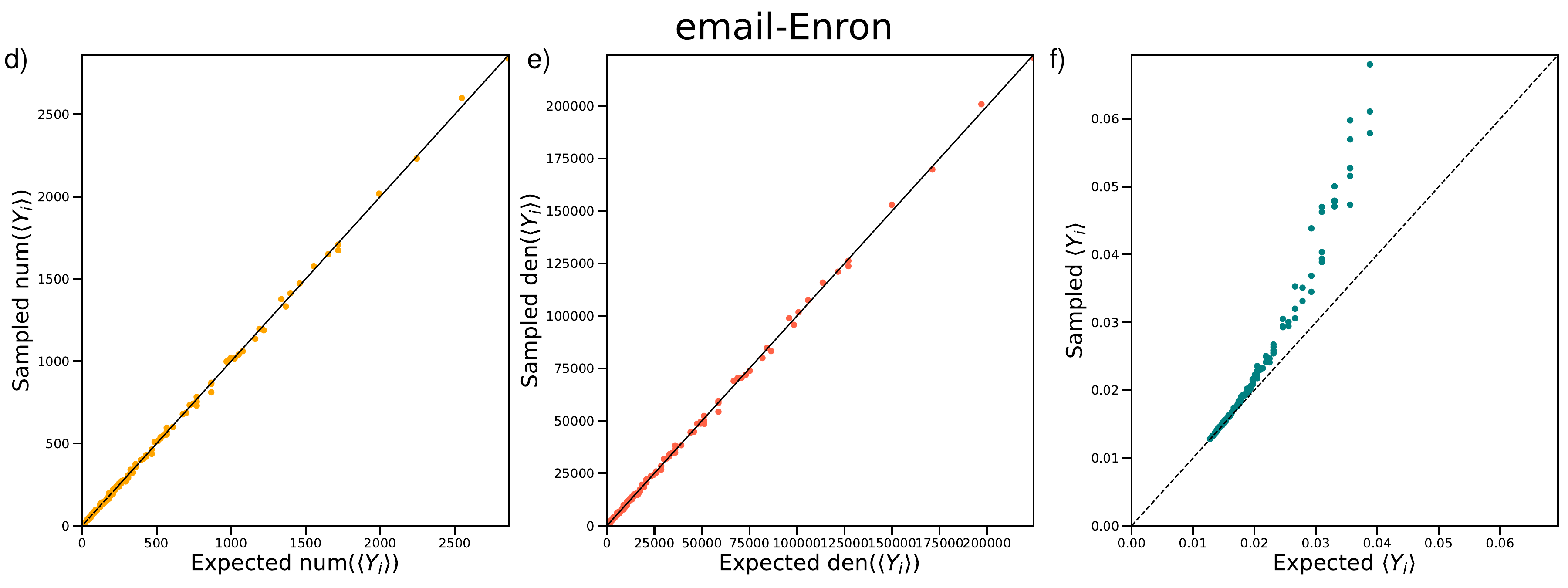}
\includegraphics[width=\textwidth]{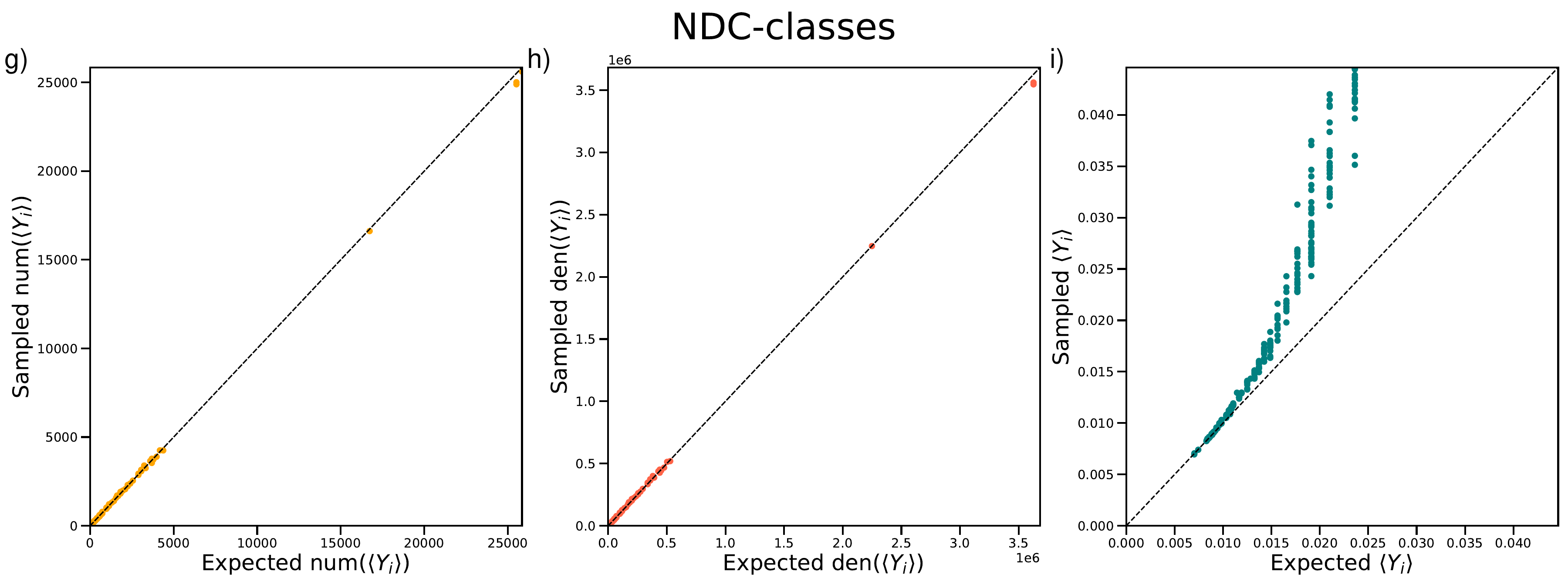}
\caption{\textbf{The limits of analytical estimates of the disparity ratio under HCM.} Panels a, d and g: scatter plots between the analytical and the sample estimates of the numerator of the disparity ratio, under the `full' HCM. Panels b, e and h: scatter plots between the analytical and the sample estimates of the denominator of the disparity ratio, under the `full' HCM. Panels c, f and i: scatter plots between the expected values $\{\langle Y_i\rangle_\text{HCM}\}$, computed as in Eq.~(\ref{eq:expdis}), and the ones obtained by explicitly sampling the ensemble induced by the `full' HCM. As analytical and sampled values of numerators and denominators nicely agree, the result concerning $\{\langle Y_i\rangle\}$ must be imputed to the correlation between the two, disregarded by Eq.~(\ref{eq:expdis}).}
\label{figB3}
\end{figure}

Let us, now, consider the strength sequence of matrix $\mathbf{W}$, for which the following results hold true

\begin{align}
\langle\sigma_i\rangle&=\sum_{\alpha=1}^Lp_{i\alpha}(\langle h_{\alpha}\rangle-p_{i\alpha}),\\
\text{Var}[\sigma_i]&=\sum_{\substack{j=1\\j\neq i}}^N\text{Var}[w_{ij}]+2\sum_{\substack{j<k\\j,k\neq i}}\text{Cov}[w_{ij},w_{ik}]\nonumber\\
&=\sum_{\substack{j=1\\j\neq i}}^N\text{Var}[w_{ij}]+2\sum_{\substack{j<k\\j,k\neq i}}[\langle w_{ij}w_{ik}\rangle-\langle w_{ij}\rangle\langle w_{ik}\rangle].
\end{align}

As a consequence, the expected value of the disparity ratio reads

\begin{align}\label{eq:expdis}
\langle Y_i\rangle&\simeq\sum_{\substack{j=1\\j\neq i}}^N\frac{\langle w_{ij}^2\rangle}{\langle\sigma_i^2\rangle}=\sum_{\substack{j=1\\j\neq i}}^N\frac{\langle w_{ij}\rangle^2+\text{Var}[w_{ij}]}{\langle\sigma_i\rangle^2+\text{Var}[\sigma_i]}
\end{align}
where the $\simeq$ symbol is understood to approximate the expected value of a ratio as the ratio of expected values. This approximation just represents the first order of the Taylor expansion

\begin{equation}\label{eq:Taylorfrac}
\text{E}\left[\frac{X}{Y}\right]\simeq\frac{\text{E}[X]}{\text{E}[Y]}-\frac{\text{Cov}[X,Y]}{\text{E}[Y]^2}+\frac{\text{E}[X]}{\text{E}[Y]^3}\text{Var}[X];
\end{equation}
due to the correlation between numerator and denominator, the second and the third terms of Eq.~(\ref{eq:Taylorfrac}) may not be negligible: as Fig.~\ref{figB3} shows, while Eq.~(\ref{eq:expdis}) holds true for the \texttt{contact-primary-school} data set, it does not for the others.

\clearpage

\section{Expected value of the clique motif eigenvector centrality}\label{app:cec}

\begin{figure}[t!]
\includegraphics[width=\columnwidth]{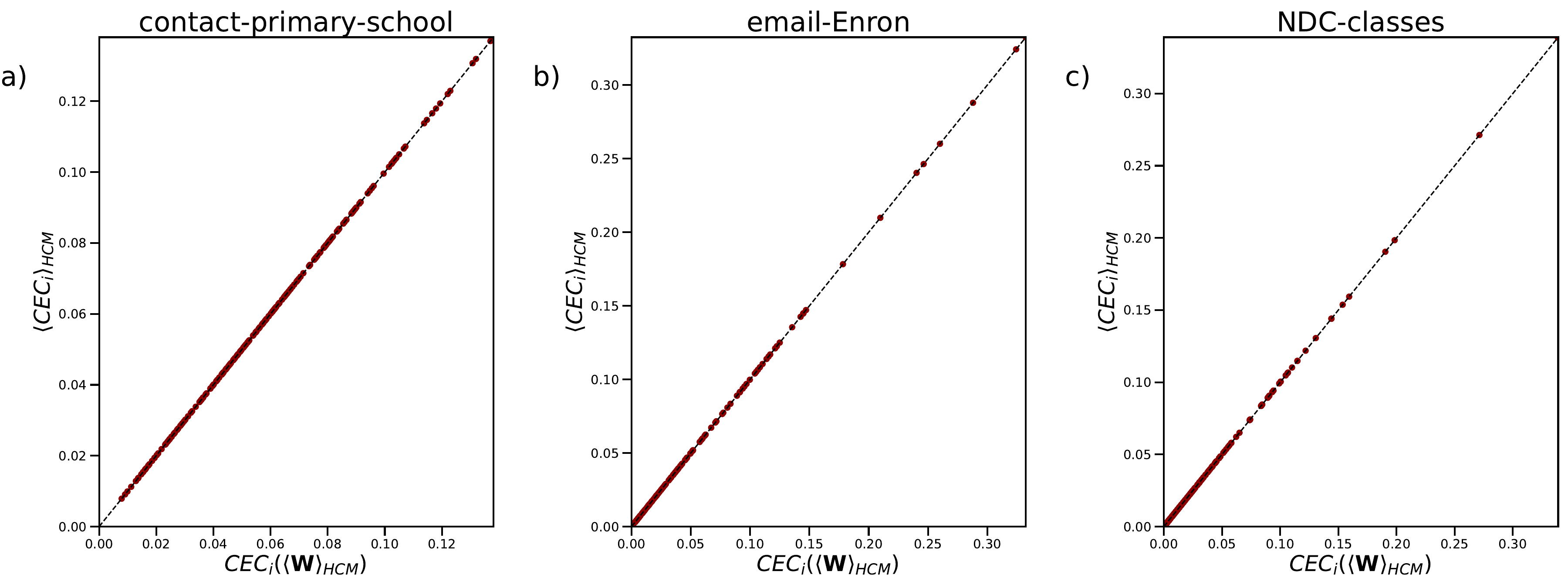}
\caption{\textbf{Comparison between the ensemble average and the analytical estimate of CEC.} Scatter plot between the ensemble average (of each component) of the Perron-Frobenius eigenvector of a given, empirical matrix $\mathbf{W}^*$ and the numerical value (of each component) of the Perron-Frobenius eigenvector of the corresponding, expected matrix $\langle\mathbf{W}\rangle_\text{HCM}$ for \texttt{contact-primary-school}, \texttt{email-Enron} and \texttt{NDC-classes} datasets (respectively, panels a, b and c). The extremely good agreement suggests a faster way to calculate the CEC that avoids to explicitly sample the ensemble induced by a null model.}
\label{fig11}
\end{figure}

The clique motif eigenvector centrality (CEC) has been proposed in~\cite{Benson2019}: its $i$-th entry, $\text{CEC}_i$, is nothing else that the corresponding entry of the Perron-Frobenius eigenvector of $\mathbf{W}$.

In order to evaluate the expected value of the CEC, we have explicitly sampled the ensemble of incidence matrices induced by the HCM; then, we have calculated the matrix $\tilde{\mathbf{W}}$ induced by each $\tilde{\mathbf{I}}$, according to Eq.~(59) of the main text. Afterwards, we have calculated the Perron-Frobenius eigenvector of each `projected' matrix and taken their average in an entry-wise fashion.

Remarkably, we found that the aforementioned ensemble average basically coincides with the Perron-Frobenius eigenvector of the ensemble average of $\mathbf{W}$ itself, i.e. $\langle\mathbf{W}\rangle_\text{HCM}$, as Fig.~\ref{fig11} shows.

\clearpage

\section{Expected value of the entries of the confusion matrix}\label{app:conf}

\begin{table}[htb!]
\captionsetup{width=\textwidth}
\caption{\textbf{Confusion matrix and density of $1$s for the three real-world hypergraphs considered in the present paper.}}
\label{tab:confusion}
\centering
\begin{tabular}{l|c|c|c|c|c}
\hline
Data set & $\langle\text{TPR}\rangle$ & $\langle\text{SPC}\rangle$ & $\langle\text{PPV}\rangle$ & $\langle\text{ACC}\rangle$ & $\rho$ \\
\hline
\hline
\texttt{contact-primary-school} & $3.57\times10^{-8}$ & 0.991 & $3.57\times10^{-8}$ & 0.983 & $8.66\times10^{-3}$ \\
\texttt{email-Enron} & $8.15\times10^{-8}$ & 0.982 & $8.15\times10^{-8}$ & 0.966 & $1.72\times10^{-2}$ \\
\texttt{NDC-classes} & $2.40\times10^{-10}$ & 0.997 & $2.40\times10^{-10}$ & 0.995 & $2.71\times10^{-3}$ \\
\hline
\end{tabular}
\end{table}

Before defining the entries of the confusion matrix, let us call $\tilde{\mathbf{I}}$ the generic member of the ensemble of incidence matrices induced by one of the benchmarks considered in the present paper: the expected value of any quantity which is a function of $\tilde{\mathbf{I}}$ is readily calculated by averaging it over the ensemble itself.\\

The true positive rate (TPR) is defined as the percentage of $1$s correctly recovered by a given model and reads

\begin{equation}
\text{TPR}=\frac{\text{TP}}{T}=\sum_{i=1}^N\sum_{\alpha=1}^L\frac{I_{i\alpha}\tilde{I}_{i\alpha}}{T}\Longrightarrow\langle\text{TPR}\rangle=\frac{\langle\text{TP}\rangle}{T}=\sum_{i=1}^N\sum_{\alpha=1}^L\frac{I_{i\alpha}p_{i\alpha}}{T};
\end{equation}
the specificity (SPC) is defined as the percentage of $0$s correctly recovered by a given model and reads

\begin{align}
\text{SPC}&=\frac{\text{TN}}{NL-T}=\sum_{i=1}^N\sum_{\alpha=1}^L\frac{(1-I_{i\alpha})(1-\tilde{I}_{i\alpha})}{NL-T}\Longrightarrow\nonumber\\
\langle\text{SPC}\rangle&=\frac{\langle\text{TN}\rangle}{NL-T}=\sum_{i=1}^N\sum_{\alpha=1}^L\frac{(1-I_{i\alpha})(1-p_{i\alpha})}{NL-T}
\end{align}
the positive predictive value (PPV) is defined as the percentage of $1$s correctly recovered by a given model with respect to the total number of predicted of $1$s and reads

\begin{equation}
\text{PPV}=\sum_{i=1}^N\sum_{\alpha=1}^L\frac{I_{i\alpha}\tilde{I}_{i\alpha}}{\tilde{T}}\Longrightarrow\langle\text{PPV}\rangle=\frac{\langle\text{TP}\rangle}{\langle T\rangle}=\sum_{i=1}^N\sum_{\alpha=1}^L\frac{I_{i\alpha}p_{i\alpha}}{\langle T\rangle};
\end{equation}
lastly, the accuracy (ACC) measures the overall performance of a given reconstruction method in correctly placing both $1$s and $0$s and reads 

\begin{equation}
\text{ACC}=\frac{\text{TP}+\text{TN}}{NL}\Longrightarrow\langle\text{ACC}\rangle=\frac{\langle\text{TP}\rangle+\langle\text{TN}\rangle}{NL}.
\end{equation}

As stressed in the main text, results on the confusion matrix of a number of real-world hypergraphs reveal that the large sparsity of the latter makes it difficult to reproduce the TPR and the PPV (see Table~\ref{tab:confusion}); on the other hand, the capability of the HCM to reproduce the density of $1$s makes it capable of recovering the density of $0$s as well, thus ensuring the overall accuracy of the model to be quite large.

\clearpage

\section{Comparing the RHM and the HCM performance}\label{app:r2}

\begin{table}[t!]
\caption{\textbf{The $R^2$ index allows us to compare the performance of the RHM with that of the HCM in reproducing the patterns of our real-world hypergraphs.} While both models perform worse than the baseline model, the HCM is, overall, more accurate than the RHM in reproducing all quantities we have considered, the only exception being $\sigma$: therefore, just constraining the total number of $1$s of a given incidence matrix is, in general, not enough to achieve an accurate description of the hypergraph it represents - although constraining the degree and the hyperdegree sequences may be not enough as well.}
\label{tab:R2}
\centering
\begin{tabular}{l|c|c|c|c|c|c}
\hline
Data set & $R^2_\text{HCM}(\sigma)$ & $R^2_\text{HCM}(Y)$ & $R^2_\text{HCM}(\text{CEC})$ & $R^2_\text{RHM}(\sigma)$ & $R^2_\text{RHM}(Y)$ & $R^2_\text{RHM}(\text{CEC})$ \\
\hline
\hline
\texttt{contact-primary-school} & -2.9697 & -1.7248 & -0.1489 & -2.7349 & -1.8189 & -0.8478 \\
\texttt{email-Enron} & -0.1344 & -0.2591 & -0.0519 & -0.0131 & -0.3462 & -0.1705 \\ 
\texttt{NDC-classes} & -0.0044 & -0.1107 & -0.0482 & -0.0008 & -0.1222 & -0.0183 \\ 
\hline
\end{tabular}
\end{table}

Let us, now, inspect the performance of the RHM and that of the HCM in reproducing the patterns of our real-world hypergraphs. To this aim, let us consider the coefficient of determination that, in our case, can be re-defined as

\begin{equation}
R^2_\text{null}(\mathbf{X})=1-\frac{\sum_{i=1}^N\left(X_i-\langle X_i\rangle_\text{null}\right)^2}{\sum_{i=1}^N\left(X_i-\bar{X}\right)^2};
\end{equation}
it measures the goodness of a null model in reproducing empirical data when compared to a baseline model whose only prediction reads $\bar{X}$, i.e. the arithmetic mean of the empirical data themselves. While such a model is characterised by $R^2=0$, a model that is capable of matching the observed values exactly is characterised by $R^2=1$; models returning predictions whose discrepancies are larger than the discrepancies accompanying the predictions of the baseline model are characterised by $R^2<0$.

As Table~\ref{tab:R2} shows, the HCM is often more accurate than the RHM in reproducing the quantities we have considered. More precisely, the RHM is found to steadily perform worse than it when the CEC is considered and to perform quite similar to it when the $Y$ is considered; the opposite tendency is observed when the $\sigma$ is considered. Overall, these results confirm that just constraining the total number of $1$s of a given incidence matrix is not enough to achieve an accurate description of the corresponding hypergraph but also suggest that constraining the degree and the hyperdegree sequences may be not enough as well - as we have already noticed when commenting the highly non-trivial degree of self-organisation of the real-world hypergraphs (whose trends are) depicted in Fig.~9 of the main text.

\end{document}